\documentclass[review]{elsarticle}

\usepackage{hyperref}
\usepackage{amssymb,amsthm}
\usepackage{graphicx,psfrag,epsf}
\usepackage{amsmath,amsfonts,amssymb,amsthm}
\usepackage{graphicx,psfrag,epsf}
\usepackage{enumerate}
\usepackage{algorithm,algorithmicx,algpseudocode}
\usepackage{bm}
\usepackage{booktabs}
\usepackage{multirow} 
\usepackage{float}
\usepackage{amsfonts}
\usepackage{algcompatible,amsmath}
\usepackage{caption}
\usepackage[labelformat=simple]{subcaption}
\usepackage{afterpage}

\usepackage{dashrule}

\usepackage{amssymb}
\DeclareMathOperator{\Tr}{Tr}
\DeclareMathOperator{\Var}{Var}

\DeclareMathOperator{\Ex}{E}

\DeclareMathOperator*{\argmin}{argmin}  

\DeclareMathOperator{\Span}{span}
\DeclareMathOperator{\sign}{sign}
\DeclareMathOperator*{\vect}{vec}

\theoremstyle{definition}
\newtheorem{remark}{Remark}

\theoremstyle{theorem}
\newtheorem{theorem}{Theorem}
\algnewcommand\TO{\item[\textbf{to}]}
\theoremstyle{lemma}

\theoremstyle{plain}
\newtheorem{assumption}{C}
\usepackage{tikz}
\DeclareRobustCommand\full  {\tikz[baseline=-0.6ex]\draw[thick] (0,0)--(0.5,0);}

\DeclareRobustCommand\dashed{\hdashrule[0.5ex][x]{0.7cm}{1pt}{0.6mm}}
\DeclareRobustCommand\chain {\tikz[baseline=-0.6ex]\draw[thick,dash pattern={on 4.5pt off 2pt on 1pt off 2pt}] (0,0)--(0.6,0);}

\journal{}
\algrenewcommand{\algorithmiccomment}[1]{\hfill$\blacktriangleright$ #1}

\usepackage{numcompress}\biboptions{authoryear}

\begin{document}

\begin{frontmatter}

\title{Smooth LASSO Estimator for the Function-on-Function \\ Linear Regression Model}

%

\author[mymainaddress]{Fabio Centofanti}
\author[mymainaddress2]{Matteo Fontana}
\author[mymainaddress]{Antonio Lepore}
\author[mymainaddress2]{Simone Vantini\corref{mycorrespondingauthor}}
\cortext[mycorrespondingauthor]{Corresponding author}\ead{simone.vantini@polimi.it}
\address[mymainaddress]{Department of Industrial Engineering, University of Naples Federico II, Piazzale Tecchio 80, 80125, Naples, Italy}
\address[mymainaddress2]{MOX - Modelling and Scientific Computing, Department of Mathematics, Politecnico di Milano, Piazza Leonardo da Vinci 32, 20133, Milan, Italy}

\begin{abstract}
A new estimator, named  S-LASSO, is proposed for the coefficient function of the  Function-on-Function  linear regression model. The S-LASSO estimator is shown to be able  to increase the interpretability of the model, by better locating regions where the coefficient function is zero,  and to smoothly estimate non-zero values of the coefficient function. The sparsity of the estimator is ensured by a \textit{functional LASSO penalty}, which pointwise shrinks toward zero  the coefficient function, while the smoothness is provided by two roughness penalties that penalize the curvature of the final estimator.
The resulting estimator  is  proved to be estimation and pointwise sign consistent. Via an extensive Monte Carlo simulation study, the estimation and predictive performance of the S-LASSO estimator are shown  to be better than (or at worst comparable with)  competing estimators already presented in the literature before.
Practical advantages  of the S-LASSO estimator are illustrated through the analysis of   the  \textit{Canadian weather},  \textit{Swedish mortality} and \textit{ship CO\textsubscript{2} emission data}. 
The S-LASSO method is implemented in the  \textsf{R} package \textsf{slasso}, openly available online on CRAN.

\end{abstract}


\end{frontmatter}
\section{Introduction}
Functional linear regression (FLR) is the generalization of the classical multivariate regression  to the context of the functional data analysis (FDA) (e.g. \cite{ramsay2005functional,horvath2012inference,hsing2015theoretical,kokoszka2017introduction}), where either the  predictor or the response or both have a functional form.
In particular,  we study  the Function-on-Function (FoF) linear regression model,  where both the predictor and the response variable are functions and each value of the latter, for any domain  point, depends on the full trajectory of the former.
The model is as follows
\begin{equation}
\label{eq_lm}
Y_i\left(t\right)=\int_{\mathcal{S}}X_{i}\left(s\right)\beta\left(s,t\right)ds+\varepsilon_{i}\left(t\right) \quad t\in\mathcal{T},
\end{equation}
for $i=1,\dots,n$. 
The pairs $\left(X_{i}, Y_i\right)$ are independent realizations of the predictor $X$ and  the response $Y$, which are assumed to be smooth random process with realizations in $L^2 (\mathcal{S})$ and $L^2 (\mathcal{T})$, i.e., the
Hilbert spaces of square integrable functions defined on the compact sets $\mathcal{S}$ and $\mathcal{T}$, respectively. Without loss of generality, the latter are also assumed  with functional mean equal to zero. 
The functions $\varepsilon_{i}$ are zero-mean random errors, independent of $X_{i}$, and have autocovariance function $K\left(t_1,t_2\right)$, $t_1$ and $t_2 \in \mathcal{T}$. The function   $\beta$ is smooth  in $L^2 (\mathcal{S}\times\mathcal{T})$, i.e., the Hilbert space of bivariate square integrable functions defined on the compact set  $\mathcal{S}\times\mathcal{T}$, and is hereinafter referred to as  \textit{coefficient function}.

FLR analysis  is a hot topic in the FDA literature. A comprehensive  review of the main results is provided by \cite{morris2015functional} as well as by
\cite{ramsay2005functional, horvath2012inference} and \cite{cuevas2014partial} who give worthwhile modern perspectives. Although the research efforts have been focused mainly on the case where either the predictor or the response have functional form \citep{cardot2003spline,li2007rates,hall2007methodology,abramowicz2018scandinavian}, the interest in the FoF linear regression has increased  in the very last years. 
In particular,	\cite{besse1996approximation} developed a spline based approach to estimate the coefficient function $\beta$, while   \cite{ramsay2005functional} proposed an estimator  assumed to be in a finite dimension tensor space spanned by two basis sets  and where regularization is achieved by  either  truncation or roughness penalties.
\cite{yao2005regression} built up an estimation method based on the principal component decomposition of the autovariance function of both the predictor and the response based on the \textit{principal analysis by conditional expectation } (PACE) method \citep{yao2005functional}. This estimator was extended by \cite{chiou2014multivariate} to the case of multivariate functional responses and predictors.
A general framework for the estimation of the coefficient function was proposed by \cite{ivanescu2015penalized} by means of the mixed model representation of the penalized regression. An extension of the ridge regression \citep{trevor2009elements} to the FoF linear regression  with an application to the Italian gas market was presented in \cite{canale2016constrained}.
To take into account the case when the errors $\varepsilon_{i}$ are correlated, in \cite{scheipl2015functional} the authors developed a general framework for additive mixed models by extending the work of \cite{ivanescu2015penalized}.

Analogously to the classical multivariate setting, in  Equation \eqref{eq_lm} the functional predictor $X$ contributes linearly   to the response $Y$ through the coefficient function $ \beta $, which works as a continuous  weight function.  Trivially note that, in the domain regions over which  $ \beta $ is equal to zero (if any),  changes in the functional predictor  $X$ do not  affect  the conditional value of $ Y $.
Because of the infinite dimensional nature of the FLR problem, coefficient functions that are sparse, i.e.,  zero valued over large parts of domain, arise very often in real applications. When the aim of the analysis is descriptive, that is the interest relies on understanding the relationship between $X$ and $Y$, rather than predictive only,  methods  that are able to capture the sparse nature of the coefficient function may be of great practical interest. These methods are referred to as \textit{sparse} or \textit{interpretable}   because they allow better interpreting the effects of the predictor on the response and reveal the  sparse nature  of $ \beta $.
On the contrary, the interpretation of the relationship between $ X $ and $ Y $ is often cumbersome for methods that do not focus on the  sparsity of the coefficient function. 
In particular, here the  interpretability of the model in Equation \eqref{eq_lm} is ultimately related to the knowledge of the  parts of the domain $\mathcal{S}\times\mathcal{T}$ where  $\beta$ is equal or different to zero, which are hereinafter referred to as \textit{null } and \textit{non-null regions},  respectively. 

Few works address the issue of the interpretability in FLR.
In the scalar-on-function setting,  \cite{james2009functional}  proposed the FLiRTI (\textit{Functional Linear Regression That's Interpretable}) estimator  that is able to recover the sparseness of the coefficient function, by imposing  $L_{1}$-penalties on the coefficient function itself and its first two derivatives.  \cite{zhou2013functional} introduced an estimator obtained in two stages  where the initial estimate is obtained by means of a Dantzig  selector \citep{candes2007dantzig} refined via the group \textit{Smoothly Clipped Absolute Deviation} (SCAD) penalty \citep{fan2001variable}.
The most recent work that addresses the issue of interpretability is that of \cite{lin2017locally}, who proposed a \textit{Smooth and Locally Sparse} (SLoS) estimator  of the coefficient function based on the combination of the smoothing spline method with the functional SCAD penalty.

However, to the best of the author knowledge, no effort has been made in the literature to obtain an interpretable estimator for the FoF linear regression model.	
In this work, we try to fill this gap by developing an interpretable estimator of the coefficient function $\beta$,  named S-LASSO  (Smooth plus LASSO) that is \textit{locally sparse} (i.e., is zero  on the null region) and, at the same time, \textit{smooth} on the non-null region.
The property of sparseness of the S-LASSO estimator is provided by  a \textit{functional LASSO penalty} (FLP), which is  the functional generalization of the classical  Least Absolute Shrinkage and Selection Operator (LASSO) \citep{tibshirani1996regression}.
Whereas, two roughness penalties, introduced in the objective function,  ensure  smoothness of the estimator on the non-null region.
From a computational point of view,  the S-LASSO estimator is obtained as the solution of  a single optimization problem by means of a new version of the \textit{orthant-wise limited-memory quasi-Newton} (OWL-QN) algorithm \citep{andrew2007scalable}, which is specifically designed to solve optimization problems involving  $L_{1}$ penalties.
The  method presented in this article is implemented in the  \textsf{R} package \textsf{slasso},  openly available online on CRAN.

\begin{figure}
	
	\centering
	\begin{subfigure}[b]{0.45\textwidth}
		
		\centering
		\includegraphics[width=\textwidth]{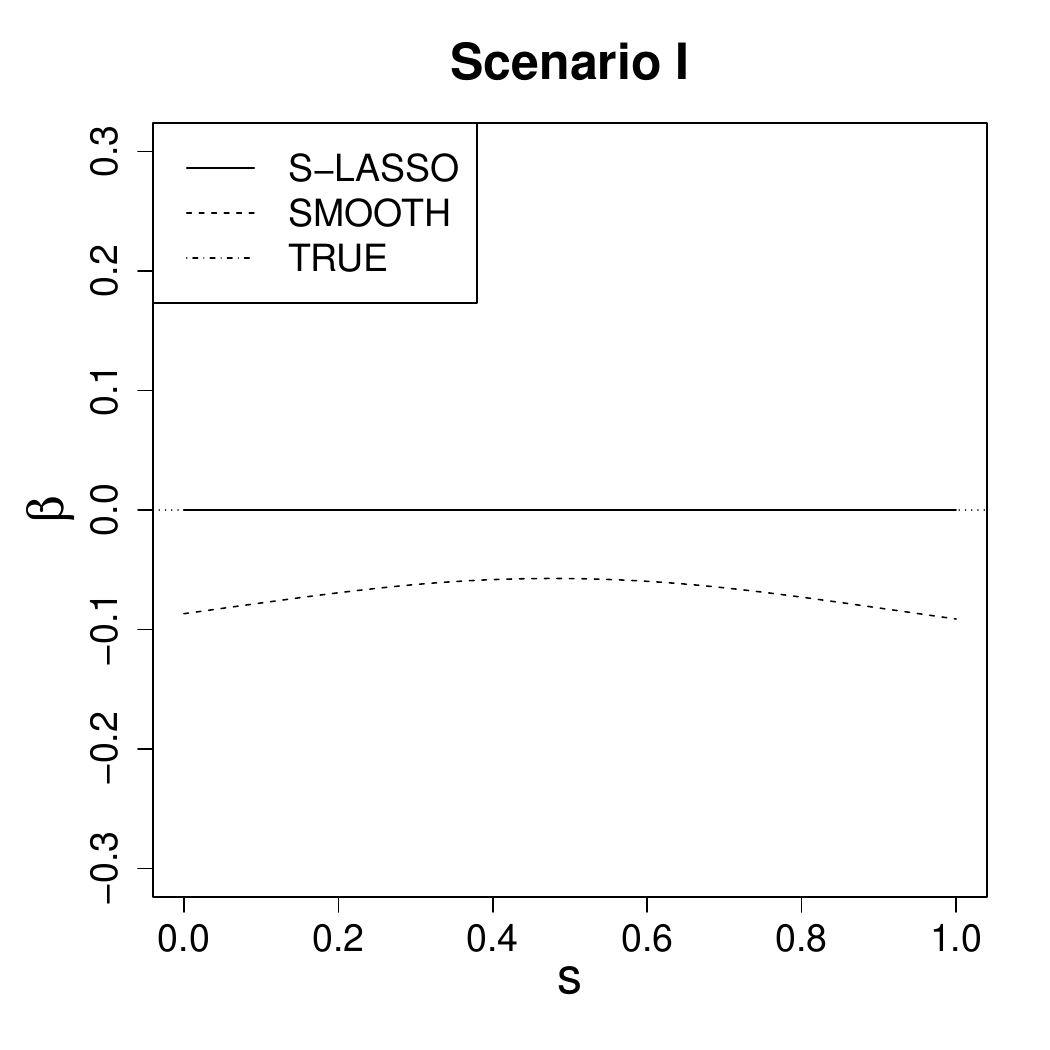}
		\vspace{-0.8cm}
		\caption{}
		
	\end{subfigure}
	\begin{subfigure}[b]{0.45\textwidth}
		
		\centering
		\includegraphics[width=\textwidth]{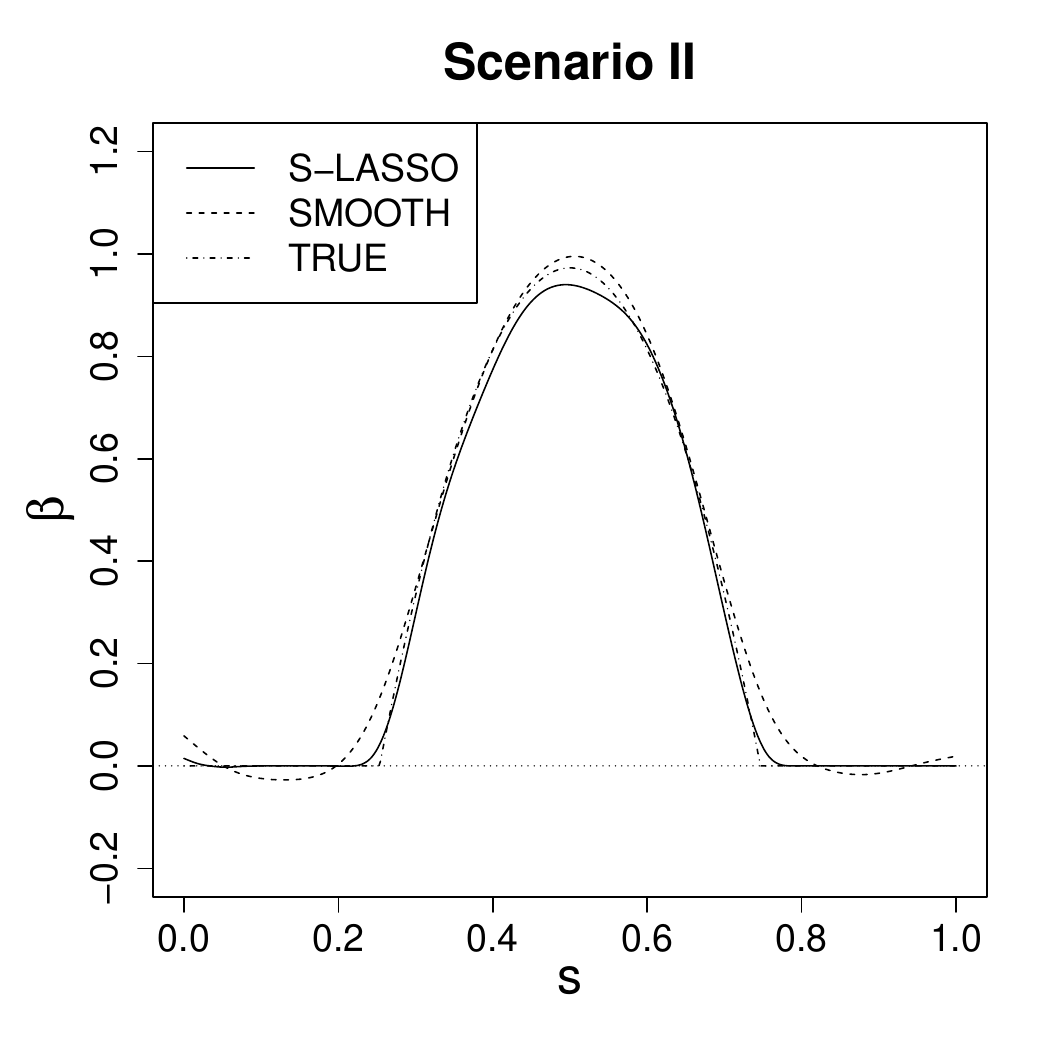}
		\vspace{-0.8cm}
		\caption{}
		
	\end{subfigure}
	\begin{subfigure}[b]{0.45\textwidth}
		
		\centering
		\includegraphics[width=\textwidth]{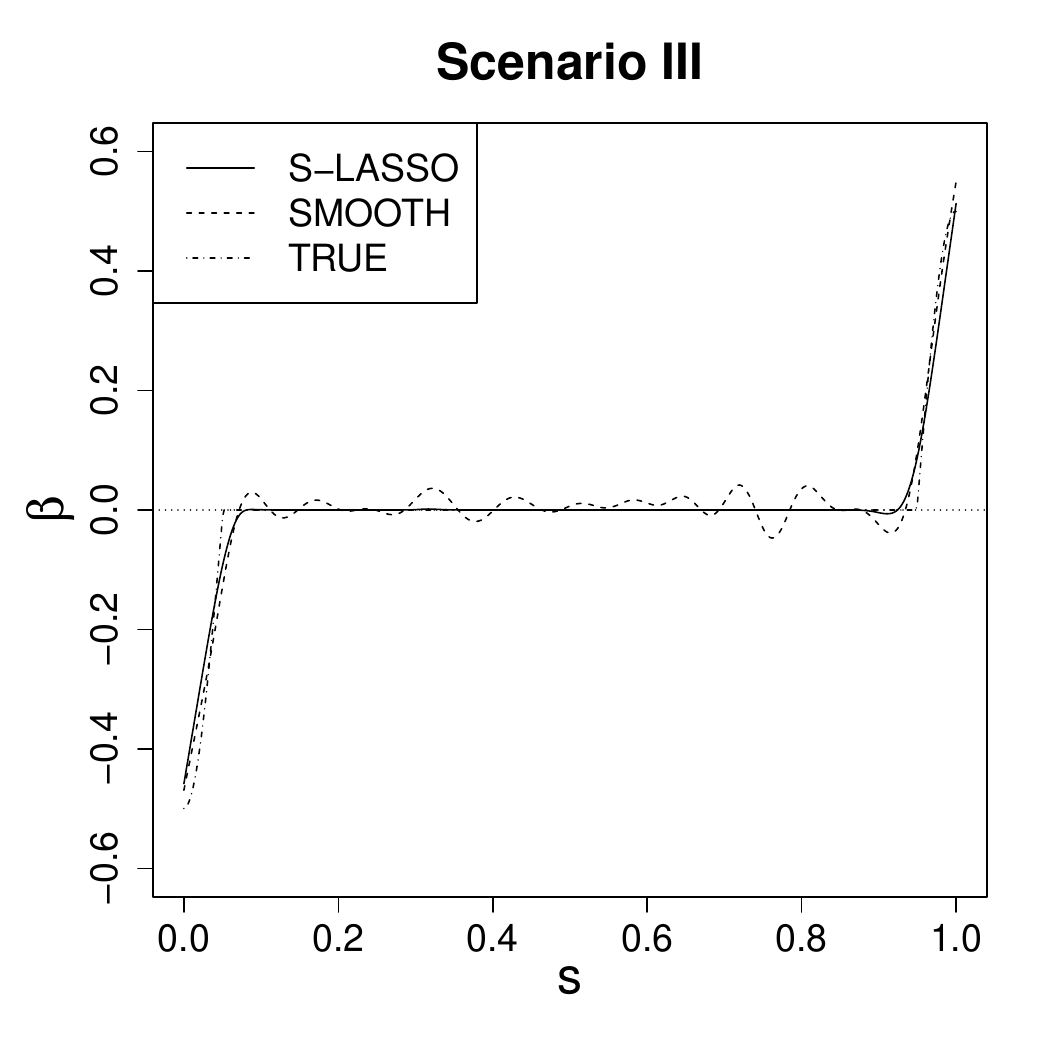}
		\vspace{-0.8cm}
		\caption{}
		
	\end{subfigure}
	\begin{subfigure}[b]{0.45\textwidth}
		\centering
		\includegraphics[width=\textwidth]{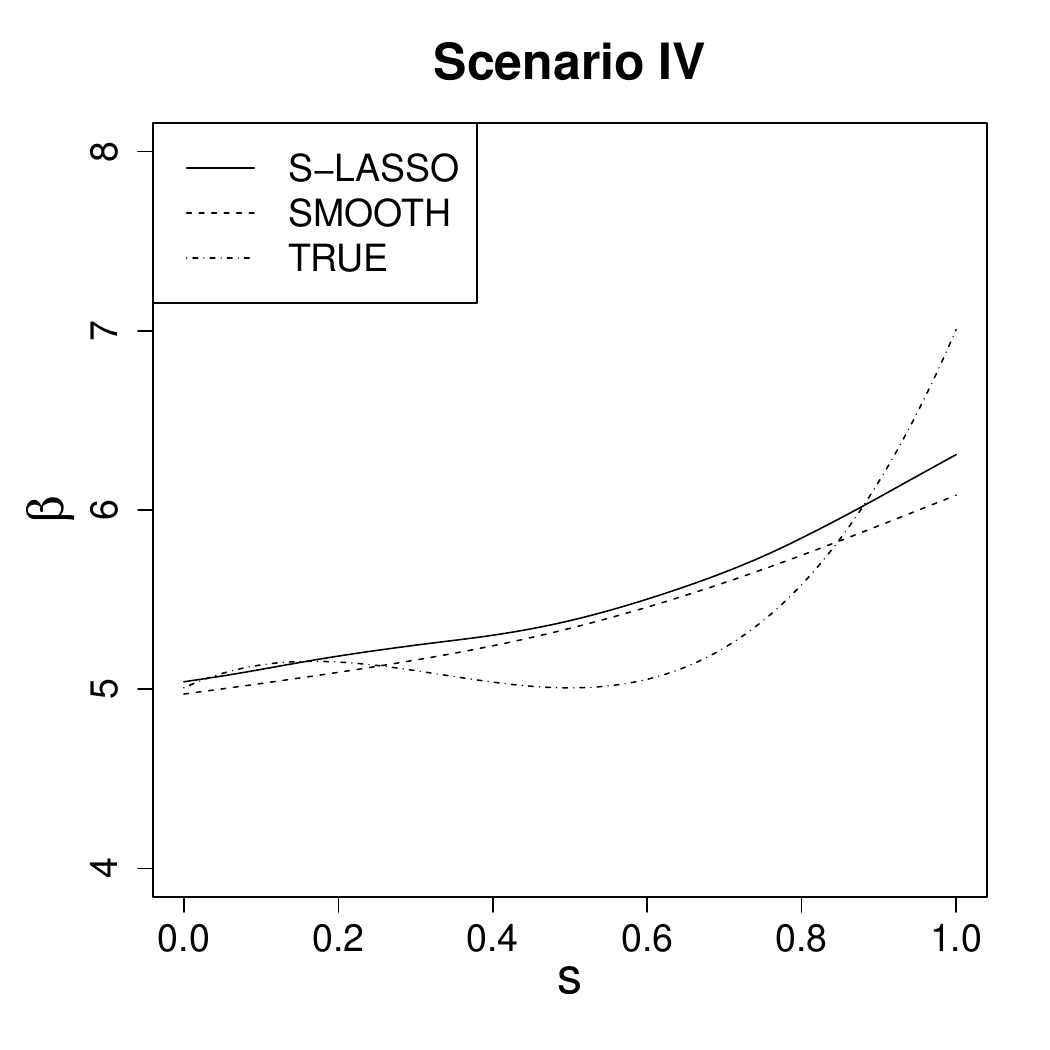}
		\vspace{-0.8cm}
		\caption{}
		
	\end{subfigure}

	\caption{The S-LASSO (\full) and SMOOTH (\dashed) estimates along with the true coefficient function, referred to as TRUE (\chain) at $t=0.5$ for Scenario I (a), Scenario II (b), Scenario III (c) and Scenario IV (d) in the simulation study detailed in Section \ref{sec_sim}. }
	\label{fig_betahat}
	
\end{figure}

To give an idea of the properties of the proposed estimator, in Figure \ref{fig_betahat} the S-LASSO estimator is applied to four different scenarios, whose data generation is detailed in Section \ref{sec_sim}.
In particular, in each plot the  S-LASSO estimate, the true coefficient function, and the smoothing spline estimate proposed by \cite{ramsay2005functional}, referred to as SMOOTH, are shown for $ t=0.5 $.   
For Scenario I, the true coefficient function is zero all over the domain, which means that the predictor $X$ is independent of the response. It is clear from Figure \ref{fig_betahat}(a) that the S-LASSO estimate successfully recovers the sparseness of $ \beta $, the same cannot be said  for the SMOOTH estimate, which is different from zero for all values of $s$.
Figure \ref{fig_betahat}(b) and Figure \ref{fig_betahat}(c) show the coefficient function estimates for Scenario II and Scenario III, where  $ \beta $ is zero on the edge and in the central part of the domain, respectively.
Also in this case the proposed method provides an estimate which is sparse on the null region and smooth on the non-null one. 
Indeed, for Scenario II, the S-LASSO estimate is zero for $ s\in\left[0,0.2\right] $ and for $ s\in\left[0.8,1\right] $, whereas, it well resembles the true coefficient function in the central part of the domain. In Figure  \ref{fig_betahat}(c), the sparsity property of the S-LASSO estimator can be further appreciated, which, in fact, over the central domain region, i.e., for  $ s\in\left[0.1,0.9\right] $,  successfully estimates   $ \beta $.
The SMOOTH estimate  in both scenarios  is not able instead to capture the sparse nature of the coefficient function.
The last scenario in Figure \ref{fig_betahat}(d) is not favourable to the proposed estimator because the true coefficient function is not sparse. However, also in this case the S-LASSO method provides  satisfactory results.
These four examples are provided with the preliminary purpose of giving an idea about the ability of the S-LASSO estimator to recover sparsity in the coefficient function while simultaneously estimating the relationship between $ X $ and $ Y $ over the non-null region. The performance of the S-LASSO estimator will be deeply analysed in  Section \ref{sec_sim}.

%

The paper is organized as follows. In Section \ref{sec_2}, the S-LASSO estimator is presented.
In Section \ref{sec_theo}, asymptotic properties of the S-LASSO estimator  are discussed  in terms of consistency and pointwise sign consistency.
In Section \ref{sec_sim}, by means of a Monte Carlo simulation study, the S-LASSO estimator is compared, in terms of estimation error and prediction accuracy, with  competing estimators already proposed in the literature. 
In Section \ref{sec_real},  the potential of the S-LASSO estimator are demonstrated with respect of   three  datasets:  the  \textit{Canadian weather},  \textit{Swedish mortality} and \textit{ship CO\textsubscript{2} emission data}.
Proofs,  data generation procedures  in the simulation study as well as additional simulation results, and,  algorithm description are given in the Supplementary Material.

\label{sec:1}
\section{Methodology}
\label{sec_2}
In Section \ref{sec_smootspline}, we briefly describe the smoothing spline estimator. Readers who are already familiar with this approach may skip to the next subsection.
In Section \ref{sec_slasso}, the S-LASSO estimator definition is given along with details on both  computational issues and model selection.
\subsection{The smoothing spline estimator}
\label{sec_smootspline}
The smoothing spline estimator of the FoF linear regression model \citep{ramsay2005functional} is the first key component of the S-LASSO estimator.
It is based on the assumption that the coefficient function $\beta$ may be well approximated by an element in the tensor product space  generated by  two  spline function spaces, where a spline is a function defined piecewise by polynomials. Well-known basis functions for  the spline space are the B-splines.
A B-spline basis is a set of spline functions uniquely defined by an order $k$  and a non-decreasing sequence of $M+2$ knots, that we hereby assume to be equally spaced in a general domain $\mathcal{D}$. Cubic B-splines are B-splines of order $k=4$. Each B-spline function is a positive polynomial of degree $k-1$ over each subinterval defined by the knot sequence and is non-zero over no more than $k$ of these subintervals (i.e., the compact support property).
In our setting, besides the computational advantage \citep{trevor2009elements}, the compact support property is fundamental because it allows one  to link the values of $\beta$ over a given region to the B-splines with support in the same region and to discard all the B-splines that are outside that region. 
%
%
Thorough descriptions  of splines and B-splines are in \cite{de2001practical} and \cite{schumaker2007spline}.

The smoothing spline estimator \citep{ramsay2005functional} is defined as
\begin{multline}
\label{eq_smoothest}
\hat{\beta}_{S}=\argmin_{\alpha \in \mathbb{S}_{k_1,k_2,M_1,M_2}}\Big\{ \sum_{i=1}^{n}||Y_{i}-\int_{\mathcal{S}}X_{i}\left(s\right)\alpha\left(s,\cdot\right)ds||^{2}+\lambda_{s}||\mathcal{L}_{s}^{m_{s}}\alpha||^{2}+\lambda_{t}||\mathcal{L}_{t}^{m_{t}}\alpha||^{2}\Big\},
\end{multline}
where $\mathbb{S}_{k_1,k_2,M_1,M_2}$ is the tensor product space generated by the sets of B-splines of orders $k_1$ and $k_2$ associated with the  non-decreasing sequences of $M_1+2$ and  $M_2+2$ knots defined on $\mathcal{S}$ and $\mathcal{T}$, respectively. $\mathcal{L}_{s}^{m_{s}}$ and $\mathcal{L}_{t}^{m_{t}}$, with $m_s\leq
k_1-1$ and $m_t\leq
k_2-1$,  are the $m_{s}$-th and $m_{t}$-th order linear differential operators  applied to $\alpha$ with respect to the variables $s$ and $t$, respectively. The symbol  $||\cdot||$ denotes the $L^{2}$-norm corresponding to the inner product $<f,g>=\int fg$. 
The parameters $\lambda_{s},\lambda_{t}\geq 0$ are   generally referred to as \textit{roughness parameters}.
The aim of the second and third terms on the right-hand side of Equation \eqref{eq_smoothest} is that of penalizing features  along  $s$ and $t$ directions.
A common practice, when dealing with cubic splines, is to  choose  $m_{s}=2$ and $m_{t}=2$, which results in the  penalization of the curvature of the final estimator.
When $\lambda_s=\lambda_t=0$, the wiggliness of the estimator is not penalized and the resulting estimator is the one that minimizes the sum of squared errors. On the contrary, as $\lambda_s\rightarrow\infty$ and $\lambda_t\rightarrow\infty$, $\hat{\beta}_{S}$ converges to a bivariate polynomial with  degree equal to $|\max\left(m_s,m_t\right)-1|$.
However, there is no guarantee that $\hat{\beta}_{S}$ is a sparse estimator, i.e., it is exactly equal to zero in some part of the domain $\mathcal{S\times T}$.

\subsection{The S-LASSO Estimator}
\label{sec_slasso}
Based on the smoothing spline estimator of  Equation \eqref{eq_smoothest}, the S-LASSO estimator is defined as follows
\begin{equation}
\label{eq_slasso}
\hat{\beta}_{SL}=\argmin_{\alpha \in \mathbb{S}_{k_1,k_2,M_1,M_2}}\Big\{ \sum_{i=1}^{n}||Y_{i}-\int_{\mathcal{S}}X_{i}\left(s\right)\alpha\left(s,\cdot\right)ds||^{2}+\lambda_{s}||\mathcal{L}_{s}^{m_{s}}\alpha||^{2}+\lambda_{t}||\mathcal{L}_{t}^{m_{t}}\alpha||^{2}+\lambda_L\int_{\mathcal{S}}\int_{\mathcal{T}}|\alpha\left(s,t\right)|dsdt\Big\}.
\end{equation}
The  last term in the right-hand side of Equation \eqref{eq_slasso} is the  extension of the LASSO penalty \citep{tibshirani1996regression} to  the FoF linear regression setting, which is referred to as \textit{functional LASSO penalty} (FLP). In particular, by starting from the multivariate LASSO penalty, the FLP is built by substituting summation with integration and by taking into account the continuity of the domain.
The FLP is able  to pointwise shrink  the value of the coefficient function for each $ s,t \in\mathcal{S}\times\mathcal{T} $.   Due to the property of the absolute value function of being singular at zero,  some of these values are shrunken exactly to zero. Thus, the FLP allows  $\hat{\beta}_{SL}$ to be exactly zero over the null region.
The constant $\lambda_{L}\geq0$ is usually called \textit{regularization parameter} and controls the degree of shrinkage towards zero of the FLP. The larger this value, the larger the shrinkage effect and the domain portion where   the resulting estimator is  zero.
Moreover, the  FLP is expected to be able to improve the prediction accuracy (in terms of expected mean square error) by introducing a bias in the final estimator.

The second and third terms on the right-hand side of Equation \eqref{eq_slasso} represent the two roughness penalties introduced in Equation \eqref{eq_smoothest} to control the  smoothness of the coefficient function estimator.

It is worth noting that, in general, the estimator smoothness may be also controlled  by opportunely choosing the dimension of the space $\mathbb{S}_{k_1,k_2,M_1,M_2}$, that is, by fixing $k_1$ and $k_2$, and choosing $M_1$ and $M_2$ \citep{ramsay2005functional}. However, this strategy is not suitable in this case.
To obtain a sparse estimator, the dimension of the space $\mathbb{S}_{k_1,k_2,M_1,M_2}$ must be in fact as large as possible.
In this way, the value of $\beta$ in a given region is strictly related to the coefficients of the B-spline functions defined on the same part of the domain and, thus, they tend to be zero in the null region. On the contrary, when the dimension of $\mathbb{S}_{k_1,k_2,M_1,M_2}$ is small,  there is a larger  probability that some B-spline functions  have support  both in the null and non-null regions and, thus the  corresponding B-spline coefficients result different from zero. 
Therefore, we find suitable the use of the two roughness penalty terms in Equation \eqref{eq_slasso}.

To compute the S-LASSO estimator, let us consider the space $\mathbb{S}_{k_1,k_2,M_1,M_2}$   generated by the two sets of  B-splines $\bm{\psi}^{s}=\left(\psi^{s}_{1},\dots,\psi^{s}_{M_1+k_1}\right)^{T}$ and $\bm{\psi}^{t}=\left(\psi^{t}_{1},\dots,\psi^{t}_{M_2+k_2}\right)^{T}$, of order $k_1$ and $k_2$ and non-decreasing knots sequences $\Delta^{s}=\lbrace s_{0},s_{1},\dots,s_{M_1},s_{M_1+1}\rbrace$ and $\Delta^{t}=\lbrace t_{0},t_{1},\dots,t_{M_2},t_{M_2+1}\rbrace$, defined on $\mathcal{S}=\left[s_0,s_{M_1+1}\right]$ and $\mathcal{T}=\left[t_0,t_{M_2+1}\right]$, respectively.
Similarly to the standard smoothing spline estimator, by performing the minimization in Equation \eqref{eq_slasso} over $\alpha\in\mathbb{S}_{k_1,k_2,M_1,M_2}$, we  implicitly assume that $\beta$ can be suitably approximated by an element in $\mathbb{S}_{k_1,k_2,M_1,M_2}$,  that is
\begin{equation}
\beta\left(s,t\right)\approx\tilde{\beta}\left(s,t\right)\doteq\sum_{i=1}^{M_1+k_1}\sum_{j=1}^{M_2+k_2}b_{ij}\psi^{s}_{i}\left(s\right)\psi^{t}_{j}\left(t\right)=\bm{\psi}^{s}\left(s\right)^{T}\bm{B}\bm{\psi}^{t}\left(t\right) \quad s\in \mathcal{S}, t\in\mathcal{T},
\end{equation}
where  $\bm{B}=\lbrace b_{ij}\rbrace\in\mathbb{R}^{M_1+k_1\times M_2+k_2}$ and $b_{ij}$ are scalar coefficients.
So stated, the problem of estimating $\beta$ has been reduced to the estimation of the unknown coefficient matrix $\bm{B}$.
Let  $\alpha\left(s,t\right)=\bm{\psi}^{s}\left(s\right)^{T}\bm{B}_{\alpha}\bm{\psi}^{t}\left(t\right)$, $s\in \mathcal{S}, t\in\mathcal{T}$, in $\mathbb{S}_{k_1,k_2,M_1,M_2}$, where  $\bm{B}_{\alpha}=\lbrace b_{\alpha,ij}\rbrace\in\mathbb{R}^{M_1+k_1\times M_2+k_2}$. Then,   the first term of the right-hand side of Equation \eqref{eq_slasso} may be rewritten  as
\begin{equation}
\label{eq_sse}
\sum_{i=1}^{n}||Y_{i}-\int_{\mathcal{S}}X_{i}\left(s\right)\alpha\left(s,\cdot\right)ds||^{2}=\sum_{i=1}^{n}\int_{\mathcal{T}}Y_{i}\left(t\right)^{2}dt-2\Tr\left[\bm{X}\bm{B}_{\alpha}\bm{Y}^{T}\right]+\Tr\left[\bm{X}^{T}\bm{X}\bm{B}_{\alpha}\bm{W}_{t}\bm{B}_{\alpha}^{T}\right],
\end{equation}
whereas,  the roughness penalties on the left side of Equation \eqref{eq_slasso} become
\begin{equation}
\label{eq_pen}
\lambda_{s}||\mathcal{L}_{s}^{m_{s}}\alpha||^{2}=\lambda_{s}\Tr\left[ \bm{B}_{\alpha}^{T}\bm{R}_{s}\bm{B}_{\alpha}\bm{W}_{t}\right]\quad \lambda_{t}||\mathcal{L}_{t}^{m_{t}}\alpha||^{2}=\lambda_{t}\Tr\left[ \bm{B}_{\alpha}^{T}\bm{W}_{s}\bm{B}_{\alpha}\bm{R}_{t}\right],
\end{equation}
where $\bm{X}=\left(\bm{X}_{1},\dots,\bm{X}_{n}\right)^{T}$, with $\bm{X}_{i}=\int_{\mathcal{S}}X_{i}\left(s\right)\bm{\psi}^{s}\left(s\right)ds$, $\bm{Y}=\left(\bm{Y}_{1},\dots,\bm{Y}_{n}\right)^{T}$ with $\bm{Y}_{i}=\int_{\mathcal{T}}Y_{i}\left(t\right)\bm{\psi}^{t}\left(t\right)dt$ $\bm{W}_{s}=\int_{\mathcal{S}}\bm{\psi}^{s}\left(s\right)\bm{\psi}^{s}\left(s\right)^{T}ds$,  $\bm{W}_{t}=\int_{\mathcal{T}}\bm{\psi}^{t}\left(t\right)\bm{\psi}^{t}\left(t\right)^{T}dt$, $\bm{R}_{s}=\int_{\mathcal{S}}\mathcal{L}_{s}^{m_{s}}\left[\bm{\psi}^{s}\left(s\right)\right]\mathcal{L}_{s}^{m_{s}}\left[\bm{\psi}^{s}\left(s\right)\right]^{T}ds$ and $\bm{R}_{t}=\int_{\mathcal{T}}\mathcal{L}_{t}^{m_{t}}\left[\bm{\psi}^{t}\left(t\right)\right]\mathcal{L}_{t}^{m_{t}}\left[\bm{\psi}^{t}\left(t\right)\right]^{T}dt$.
The term $\Tr\left[\bm{A}\right]$ denotes the trace of a square matrix $\bm{A}$.

Standard optimization algorithms for $L_{1}$-regularized objective functions are designed for the case where the absolute value  enters the problem  as a  linear function of the parameters.
However, in the optimization problem in Equation \eqref{eq_slasso}, the FLP particularizes as follows
\begin{equation}
\lambda_{L}\int_{\mathcal{S}}\int_{\mathcal{T}}|\alpha\left(s,t\right)|dsdt=\lambda_L\int_{\mathcal{S}}\int_{\mathcal{T}}|\bm{\psi}^{s}\left(s\right)^{T}\bm{B}_{\alpha}\bm{\psi}^{t}\left(t\right)|dsdt,
\end{equation}
which is not a linear function of the absolute value of the coefficient matrix $|\bm{B}_{\alpha}|$, because the absolute value is instead applied to a linear combination of the  parameters.
Therefore, in order to be able to use optimization algorithms for $L_{1}$-regularized objective functions  by the following theorem, we  provide a practical way to approximate the FLP as a linear function of $|\bm{B}_{\alpha}|$, and thus extremely simplify the computation. 
\begin{theorem}
	\label{the_1}	
	Let $\mathbb{S}_{k_1,k_2,\Delta_{1,e}\Delta_{2,e}}=\Span\lbrace \psi_{i_1}\psi_{i_2}\rbrace_{i_1=1,i_2=1}^{M_1+k_1,M_2+k_2}$, with $\lbrace \psi_{i_j}\rbrace$  the set of B-splines of orders $k_j$  and non-decreasing evenly spaced knots sequences \\$\Delta_{j}=\lbrace x_{j,0},x_{j,1},\dots,x_{j,M_j},x_{j,M_j+1}\rbrace$ defined on the compact set $\mathcal{D}_j=\left[x_{j,0},x_{j,M_j+1}\right]$ and $\Delta_{j,e}$  the extended partitions corresponding to $\Delta_{j}$ defined as $\Delta_{j,e}=\lbrace y_{j,l}\rbrace_{l=1}^{M_j+2k_j}$ where $ y_{j,1},\dots,y_{j,k_j}=x_{j,0}$, $y_{j,1+k_j},\dots,y_{j,M_j+k_j}=x_{j,1},\dots,x_{j,M_j}$ and $y_{j,M_j+1+k_j},\dots,y_{j,M_j+2k_j}=x_{j,M_j+1}$, for $j=1,2$.
	Then, for $f\left(z_1,z_2\right)=\sum_{i_1=1}^{M_1+k_1}\sum_{i_2=1}^{M_2+k_2}c_{i_{1}i_{2}}\psi_{i_1}\left(z_1\right)\psi_{i_2}\left(z_2\right)\in \mathbb{S}_{k_1,k_2,\Delta_{1,e}\Delta_{2,e}}$, with $z_1\in\mathcal{D}_{1}$ and $z_2\in\mathcal{D}_{2}$,
	\begin{equation}
	0\leq||f||_{\ell^{1},\Delta_{1,e},\Delta_{2,e}}-||f||_{L^{1}}=O\left(\frac{1}{M_1}\right)+O\left(\frac{1}{M_2}\right),
	\end{equation}
	where 
	\begin{equation}
			\label{teh_approx}
	||f||_{\ell^{1},\Delta_{2,e},\Delta_{1,e}}=\sum_{i_1=1}^{M_1+k_1}\sum_{i_2=1}^{M_2+k_2} |c_{i_{1}i_{2}}|\frac{\left(y_{1,i_{1}+k_1}-y_{1,i_{1}}\right)\left(y_{2,i_{2}+k_2}-y_{2,i_{2}}\right)}{k_1 k_2},
	\end{equation}
	and 
	\begin{equation}
		\label{teh_eqnorm}
	||f||_{L^{1}}=\int_{\mathcal{D}_1}\int_{\mathcal{D}_2} |f\left(z_1,z_2\right)|dz_1dz_2.
	\end{equation}
\end{theorem}
The interpretation of the above theorem is quite simple. It basically says that for large values of $M_1$ and $M_2$, $||f||_{L^{1}}$ is well approximated from the top by $||f||_{\ell^{1},\Delta_{2,e},\Delta_{1,e}}$ and the approximation error tends to zero as $M_1,M_2\rightarrow\infty$.
By using this result, the FLP can be  approximated as follows
\begin{equation}
\label{eq_betapp}
\lambda_{L}\int_{\mathcal{S}}\int_{\mathcal{T}}|\alpha\left(s,t\right)|dsdt\approx\lambda_{L}\sum_{i=1}^{M_1+k_1}\sum_{j=1}^{M_2+k_2}|b_{\alpha,ij}|\frac{\left(s^{e}_{i+k_1}-s^{e}_{i}\right)\left(t^{e}_{j+k_2}-t^{e}_{j}\right)}{k_1 k_2}=\lambda_{L}\bm{w}_{s}^{T}|\bm{B}_{\alpha}|\bm{w}_{t},
\end{equation}
where $\lbrace s^{e}_i\rbrace$ and  $\lbrace t^{e}_i\rbrace$ are the extended partitions associated with $\Delta^{s}$ and $\Delta^{t}$, respectively,  $\bm{w}_{s}=\left[\frac{\left(s^{e}_{1+k_1}-s^{e}_{1}\right)}{k_1},\dots,\frac{\left(s^{e}_{M_1+2k_1}-s^{e}_{M_1+k_1}\right)}{k_1}\right]^{T}$ and $\bm{w}_{t}=\left[\frac{\left(t^{e}_{1+k_2}-t^{e}_{1}\right)}{k_2},\dots,\frac{\left(t^{e}_{M_2+2k_2}-t^{e}_{M_2+k_2}\right)}{k_2}\right]^{T}$.
Therefore, upon using the approximation    in Equation \eqref{eq_betapp}, Equation \eqref{eq_sse} and Equation \eqref{eq_pen}, the optimization problem in Equation \eqref{eq_slasso} becomes
\begin{align}
\label{eq_slassoobj}
\hat{\bm{B}}_{SL}\approx\argmin_{\bm{B}_{\alpha} \in \mathbb{R}^{\left(M_1+k_1\right)\times \left(M_2+k_2\right)}}\Big\{&\sum_{i=1}^{n}\int_{\mathcal{T}}Y_{i}\left(t\right)^{2}dt-2\Tr\left[\bm{X}\bm{B}_{\alpha}\bm{Y}^{T}\right]+\Tr\left[\bm{X}^{T}\bm{X}\bm{B}_{\alpha}\bm{W}_{t}\bm{B}_{\alpha}^{T}\right]\nonumber\\&+\lambda_{s}\Tr\left[ \bm{B}_{\alpha}^{T}\bm{R}_{s}\bm{B}_{\alpha}\bm{W}_{t}\right]+\lambda_{t}\Tr\left[ \bm{B}_{\alpha}^{T}\bm{W}_{s}\bm{B}_{\alpha}\bm{R}_{t}\right]+\lambda_{L}\bm{w}_{s}^{T}|\bm{B}_{\alpha}|\bm{w}_{t}\bigg\}.
\end{align}
Then, the coefficient $\beta$ is estimated by  $\hat{\beta}_{SL}\left(s,t\right)=\bm{\psi}^{s}\left(s\right)^{T}\hat{\bm{B}}_{SL}\bm{\psi}^{t}\left(t\right) $ for $s\in \mathcal{S}$ and  $t\in\mathcal{T}$.
Note that, in Equation \eqref{eq_slassoobj}, the FLP is approximated through a weighted linear combination of the absolute values of the coefficients which strictly resembles the multivariate LASSO penalty applied to the basis expansion coefficients, i.e. $\lambda_{L} \sum_{i=1}^{M_1+k_1}\sum_{j=1}^{M_2+k_2}|b_{\alpha,ij}| $. However, the presence of  $ \bm{w}_{s} $ and $ \bm{w}_{t} $ in the FLP approximation in Equation \eqref{eq_slassoobj} is crucial because it allows the penalty to differently shrink  coefficients among B-splines. That is, it avoids that the absolute values of coefficients corresponding to B-splines strictly localized are  weighted as the absolute value of coefficients of spreader basis in the computation of the penalty. This is the direct consequence of the fact that the proposed approximation is a better approximation of the FLP than the multivariate LASSO penalty applied to the coefficients.

\begin{remark}
	\label{rem_1}
	 Theorem 1 states that the $L_1$ norm of a  bivariate function in $\mathbb{S}_{k_1,k_2,\Delta_{1,e}\Delta_{2,e}}$  defined in Equation  \eqref{teh_eqnorm} could be well approximated by the expression in Equation  \eqref{teh_approx}. That is, the sum of the absolute values of the  basis coefficients $c_{i_{1}i_{2}}$  weighted by the size of the support of the corresponding basis functions divided by $k_1$ and $k_2$,  the order of the two sets of B-splines. As the dimension of the knot sequences goes to infinity, i.e., $M_1,M_2\rightarrow\infty$, Theorem 1 states that the approximation error tends to zero.
	From the proof of Theorem 1 in the Supplementary Material B, it is clear that smoothness properties of the function  $f$  influences the rate whereby the approximation error goes to zero, through the $\gamma$-modulus of smoothness $\omega_{(1,1)}$ in Equation (B.2). Practically speaking, this means the smoother the function $f$, the faster the approximation error goes to zero, and thus, small values of $M_1$ and $M_2$ are sufficient to ensure the validity of the approximation in Theorem 1.
	Because the coefficient function $\beta$ and its smoothness properties are usually unknown the appropriate choice of $M_1$ and $M_2$ is application specific. A sensitivity  analysis of the results with respect to  the choice of $ M_1 $ and $ M_2 $ could be performed, as is done in Supplementary Material D for the simulation study in Section 4. 
	The S-LASSO estimator needs also the B-splines orders $k_1$ and $k_2$ to be chosen with respect to the degree of smoothness one wants to achieve. The larger the values of $k_1$ and $k_2$, the smoother the resulting estimator. A standard choice is $k_1=k_2=4$, i.e., cubic B-splines, which ensures the continuity of the second derivative of the estimated coefficient function.
\end{remark}

\begin{remark}
	\label{rem_2}
 Theorem 1 considers specific extended knot partitions  where the first  and last knots are equal to the boundaries of the compact sets $ \mathcal{D}_j $. This derives from the Curry-Schoenberger theorem (see e.g., \cite{de2001practical},  Theorem (44)) that provides guidelines to choose an appropriate knot sequence to construct a  set of B-splines  for any particular polynomial spline space defined on a univariate closed interval. Specifically, the Curry-Schoenberger theorem requires  the knot sequence to be an \textit{extended knot partition} (\cite{schumaker2007spline}, Definition 4.8). That is, being $k$  the order of the  set of B-splines,  the first $k$ and last $k$ knots should be respectively chosen smaller than or equal to lower bound and larger than or equal to the upper bound of the definition interval of the polynomial spline space.
  A convenient choice \citep{de2001practical,ramsay2005functional} is to set them exactly equal to the lower and upper bounds, respectively.  In this way, we avoid to impose  continuity condition \citep{de2001practical} at the boundaries where, thus, differentiability of the coefficient function is lost. This choice is consistent with the fact that the set of B-splines provides a valid representation of the corresponding spline space only in the definition interval, as there are generally no  information about  the behaviour of the function to be estimated outside the definition interval, where the coefficient function is not constrained to be continuous. This intuition is directly translated to  Theorem 1, where   no continuity condition is imposed at the definition interval boundaries to the bivariate functions generated by the two sets of B-splines.
\end{remark}

The  optimization problem with $L_{1}$-regularized loss in Equation \eqref{eq_slassoobj} is (i) convex, being sum or integral of convex function;  and (ii) has a unique solution given some general conditions on the matrix $\bm{W}_{t}\otimes\bm{X}^{T}\bm{X}$ (with $\otimes$ the Kronecker product). See  Section \ref{sec_theo} for further details. Unfortunately, the objective function is not differentiable in zero, and thus it has not a closed-form solution.  In view of this,   general purpose gradient-based optimization algorithms ---as for instance the \textit{L-BFGS} quasi-Newton method \citep{nocedal2006numerical}--- and  classical optimization algorithms for solving LASSO problems ---such as coordinate descent \citep{friedman2010regularization} and least-angle regression (LARS) \citep{efron2004least}--- are  not suitable.
In contrast, we found very promising a modified version of the \textit{orthant-wise limited-memory quasi-Newton} (OWL-QN) algorithm proposed by \cite{andrew2007scalable}.	
The  OWL-QN algorithm is based on the fact that the $L_1$ norm is differentiable for the set of points named \textit{orthant} in which each coordinate never changes sign, being a linear function of its argument. In each orthant, the second-order behaviour of an objective function  of the form $f\left(\bm{x}\right)=l\left(\bm{x}\right)+C||\bm{x}||_{1}$, to be minimized, is determined by $l$ alone. The function $l:\mathbb{R}^{r}\rightarrow\mathbb{R}$ is  convex, bounded below, continuously differentiable  with continuously differentiable gradient $\nabla l$, $\bm{x}=\left(x_1,\dots,x_r\right)^{T}$, $C$ is a given positive constant, and $||\cdot||_{1}$ is the usual $L_{1}$ norm. Therefore, \cite{andrew2007scalable} propose to derive a quadratic approximation of the function $l$ that is valid for  some orthant containing the current point and then to search for the minimum of the approximation, by constraining the solution in the orthant where the approximation is valid. There may be several orthants containing or adjacent to a given point. The choice of the orthant to explore is based on the \textit{pseudo-gradiant} $\diamond f\left(\bm{x}\right)$ of $f$ at $\bm{x}$, whose components are defined as 
\begin{equation}
\label{eq_pseudo}
\diamond_{i} f\left(\bm{x}\right)=\left\{\begin{array}{ll}
\frac{\partial l\left(x\right)}{\partial x_i}+C\sign\left(x_i\right)& \text{if } |x_i|>0\\
\frac{\partial l\left(x\right)}{\partial x_i}+C& \text{if } x_i=0,\frac{\partial l\left(x\right)}{\partial x_i}<-C\\
\frac{\partial l\left(x\right)}{\partial x_i}-C& \text{if } x_i=0,\frac{\partial l\left(x\right)}{\partial x_i}>C\\
0 & \text{otherwise},
\end{array}
\right.
\end{equation}
where $\sign\left(\cdot\right)$ denotes the usual  sign function.
However, the objective function of the optimization problem in Equation \eqref{eq_slassoobj} is in the form $f^{*}\left(\bm{x}\right)=l\left(\bm{x}\right)+C||\bm{D}\bm{x}||_{1}$, with $\bm{D}=\lbrace d_i\rbrace\in\mathbb{R}^{r\times r}$  a diagonal matrix of positive weights.
To take into account these weights, the OWL-QN algorithm must be implemented with a different \textit{pseudo-gradiant}  $\diamond f^{*}\left(\bm{x}\right)$ whose components are defined as 
\begin{equation}
\label{eq_pseudo2}
\diamond_{i} f^{*}\left(\bm{x}\right)=\left\{\begin{array}{ll}
\frac{\partial l\left(x\right)}{\partial x_i}+d_i C\sign\left(x_i\right)& \text{if } |x_i|>0\\
\frac{\partial l\left(x\right)}{\partial x_i}+d_i C& \text{if } x_i=0,\frac{\partial l\left(x\right)}{\partial x_i}<-C\\
\frac{\partial l\left(x\right)}{\partial x_i}-d_i C& \text{if } x_i=0,\frac{\partial l\left(x\right)}{\partial x_i}>C\\
0 & \text{otherwise}.
\end{array}
\right.
\end{equation}
A more detailed description of the  OWL-QN algorithm for  objective functions in the form $l\left(\bm{x}\right)+C||\bm{D}\bm{x}||_{1}$ is given  in the Supplementary Material A.
Note that, the optimization problem in Equation \eqref{eq_slassoobj} can be rewritten by vectorization as
\begin{align}
\label{eq_slassoobj2}
\hat{\bm{b}}_{SL}\approx\hat{\bm{b}}_{app}=\argmin_{\bm{b}_{\alpha} \in \mathbb{R}^{\left(M_1+k_1\right) \left(M_2+k_2\right)}}\Big\{&-2\vect\left(\bm{X}^{T}\bm{Y}\right)^{T}\bm{b}_{\alpha}+\bm{b}_{\alpha}^{T}\left(\bm{W}_{t}\otimes\bm{X}^{T}\bm{X}\right)\bm{b}_{\alpha}\nonumber\\
&+\lambda_{s}\bm{b}_{\alpha}^{T}\bm{L}_{wr}\bm{b}_{\alpha}+\lambda_{t}\bm{b}_{\alpha}^{T}\bm{L}_{rw}\bm{b}_{\alpha}+\lambda_{L}||\bm{W}_{st}\bm{b}_{\alpha}||_{1}\bigg\},
\end{align}
where $\hat{\bm{b}}_{SL}=\vect\left(\hat{\bm{B}}_{SL}\right)$, $\bm{L}_{rw}\doteq\left(\bm{R}_{t}\otimes \bm{W}_{s}\right)$ and   $\bm{L}_{wr}\doteq\left(\bm{W}_{t}\otimes \bm{R}_{s}\right)$, and $\bm{W}_{st}$ is the diagonal matrix whose diagonal elements are $\bm{w}_s^{T}\otimes\bm{w}_{t}^{T}$.
Moreover, for generic a matrix $\bm{A}\in \mathbb{R}^{j\times k}$, $\vect(\bm{A})$ indicates the vector of length $jk$ obtained by writing the matrix $\bm{A}$ as a vector column-wise.
Therefore, the OWL-QN with pseudo-gradient as in Equation \eqref{eq_pseudo2} can be straightforwardly applied.

In the following, we summarize all the parameters that need to be set to obtain the S-LASSO estimator.
The orders $k_1$ and $k_2$ should be chosen with respect to the degree of smoothness we want to achieve, and the  computational efforts. 
The larger the values of $k_1$ and $k_2$, the smoother the resulting estimator will be.
Following Remark \ref{rem_1}, a standard choice are cubic B-splines, with equally spaced knot sequences, i.e.,  $k_1=k_2=4$. Moreover, $M_1$ and $M_2$ should be as large as possible to ensure that  the null region is correctly captured and  the approximation in Equation \eqref{eq_betapp} is valid, with respect to the maximum computational efforts.  
Finally,   at given $k_1$, $k_2$, $M_1$, and $M_2$, the optimal values  of $\lambda_s$, $\lambda_t$ and $\lambda_{L}$ can be  selected as those  that minimize the  the estimated prediction error  function $CV\left(\lambda_s,\lambda_t,\lambda_{L}\right)$, i.e., $CV\left(\lambda_s,\lambda_t,\lambda_{L}\right)$,   over a grid of candidate values \citep{trevor2009elements}. However,  although this choice could be optimal for the prediction performance, it may affect the interpretability of the model.
Much more interpretable models, with a slight decrease in predictive performance, may in fact exist.
The $k$\textit{-standard error} rule, which is a generalization of the \textit{one-standard error} rule \citep{trevor2009elements}, may be a more reasonable choice. That is, to choose  the most sparse model whose error is no more than $k$ standard errors above the error of the best model. 
In practice, as spareness is controlled by the parameter $\lambda_{L}$, we first find the best model in terms of estimated prediction error at given $\lambda_{L}$ and then, among the selected models, we apply the  $k$-standard error rule. This rule may be particularly useful when $CV\left(\lambda_s,\lambda_t,\lambda_{L}\right)$ is flat with respect to $\lambda_{L}$. In this case, it chooses the simplest model  among those achieving similar estimated prediction errors. The value of $ k $ quantifies the trade-off between  prediction performance and model interpretability. The larger $ k $, the lower the predictive performance but the higher the model interpretability. Commonly used values o $ k $ are 0.5, 1, 2 \citep{trevor2009elements}.

\section{Theoretical Properties of the S-LASSO Estimator}
\label{sec_theo}
In this section, we provide some theoretical results on the S-LASSO estimator,  under some regularity assumptions, i.e., the estimation consistency (Theorem \ref{the_2}) and the pointwise sign consistency (Theorem \ref{the_3}) of $\hat{\beta}_{SL}$.
All proofs are in the Supplementary Material B.

The following regularity conditions are assumed.

\begin{assumption}
	\label{as_0}
	$||X||_{2}$ is almost surely bounded, i.e., $||X||_{2}\leq c< \infty$.
\end{assumption}
\begin{assumption}
	\label{as_01}
	The null space of $ K_X $, i.e., the covariance operator   of $ X_i $,  is empty where $ K_X f(s')=\int_{\mathcal{S}}E(X(s')X(s))f(s)ds $, $ s'\in \mathcal{S} $, and $E||X||_{2}^2< \infty$.
\end{assumption}

\begin{assumption}
	\label{as_1}
	$\beta$ is in the H\"{o}lder space $C^{p',\nu}\left(\mathcal{S}\times\mathcal{T}\right)$ defined as the set of functions $f$ on $\mathcal{S}\times\mathcal{T}$ having continuous partial and mixed derivatives up to order $p'$ and such that the partial and mixed derivatives of order  $p'$ are H\"{o}lder continuous, i.e., $|f^{\left(p'\right)}\left(\bm{x}_1\right)-f^{\left(p'\right)}\left(\bm{x}_2\right)|\leq c||\bm{x_1}-\bm{x}_2||^{\nu}$, for some constant $c$, integer $p'$ and $\nu\in\left[0,1\right]$, and for all $\bm{x}_1,\bm{x}_2\in \mathcal{S}\times\mathcal{T}$, where $f^{\left(p'\right)}$ is the partial and mixed derivatives of order  $p'$. Moreover, let $p\doteq p'+\nu$ such that $3/2<p\leq k_1-1$ and $3/2<p\leq k_2-1$.
	
\end{assumption}
\begin{assumption}
	\label{as_4}
	$M_{1}=o\left(n^{1/4}\right)$, $M_{2}=o\left(n^{1/4}\right)$, $M_1=\omega\left(n^{\frac{1}{2p+1}}\right)$, $M_2=\omega\left(n^{\frac{1}{2p+1}}\right)$, where $a_{n}=\omega\left(b_n\right)$ means $\frac{a_n}{b_n}\rightarrow\infty$ for $n\rightarrow\infty$,
\end{assumption}
\begin{assumption}
	\label{as_2}
	There exist two positive constants $b$ and $B$ such that
	\begin{equation}
	b\leq\Lambda_{min}\left( \bm{W}_{t}\otimes n^{-1}\bm{X}^{T}\bm{X}\right)\leq \Lambda_{max}\left( \bm{W}_{t}\otimes n^{-1}\bm{X}^{T}\bm{X}\right)\leq B,
	\end{equation}
	where $\Lambda_{min}\left(\bm{M}\right)$ and $\Lambda_{max}\left(\bm{M}\right)$ denote the minimum and maximum eigenvalues of the  matrix $\bm{M}$.
\end{assumption}
\begin{assumption}
	\label{as_2p}
	$\lambda_{s}=o\left(M_1 ^{-2m_s+1}\right)$, $\lambda_{t}=o\left(M_2 ^{-2m_t+1}\right)$.
\end{assumption}
\noindent C.\ref{as_0} and C.\ref{as_1} are the anoulogus of   (H1) and (H2) in \cite{cardot2003spline} for a bivariate regression function.
C.\ref{as_01}  ensures identifiability of the coefficient function $ \beta $ in Equation \eqref{eq_lm} \citep{cardot2003spline,prchal2007spline,scheipl2016identifiability}.
When this assumption is not verified, the results of this section are analogously  valid by considering  the unique $ \beta $ that satisfies  Equation \eqref{eq_lm}, and, for each $ t\in\mathcal{T} $, belongs to the closure of $ Im(K_X)=\lbrace K_X f :f\in L^2(\mathcal{S})\rbrace $, i.e., the image of the covariance operator of $ X_i $ \citep{cardot2003spline}.
 C.\ref{as_1} ensures that  $\beta$ is sufficiently smooth.
C.\ref{as_4} provides information on the growth rate of the number of knots $M_1$ and $M_2$, which are strictly related to the sample size $ n $.
C.\ref{as_2} is the anolugus of condition (F) of \cite{fan2004nonconcave} and assumes that the matrix $\left( \bm{W}_{t}\otimes n^{-1}\bm{X}^{T}\bm{X}\right)$ has reasonably good behaviour, whereas, C.\ref{as_2p} provides guidance  on the choice of the parameters $\lambda_s$ and  $\lambda_t$.

Theorem \ref{the_2} shows that with probability tending to one there exists a  solution of  the optimization problem in Equation \eqref{eq_slasso} that converges to  $\tilde{\beta}$, chosen  such that $||\beta-\tilde{\beta}||_{\infty}=O(M_1^{-p})+O(M_2^{-p})$.
To prove Theorem \ref{the_2}, in addition to C.\ref{as_0}-C.\ref{as_2p}, the following condition is considered.
\begin{assumption}
	\label{as_3}
	$\lambda_L=o\left(M_1^{-1}M_2^{-1}\right)$.
\end{assumption}
\noindent The first result is about the convergence rate of $\hat{\beta}_{SL}$ to $\beta$ in terms of $L_{\infty}$-norm.
\begin{theorem}
	\label{the_2}	
	Under assumptions C.\ref{as_0}-C.\ref{as_3}, there exists a unique solution $\hat{\beta}_{SL}$ of the optimization problem in Equation \eqref{eq_slasso}, such that
	\begin{equation}
	||\hat{\beta}_{SL}-\beta||_{\infty}=
	O_{p}\left(M_{1}^{1/2}M_{2}^{1/2}n^{-1/2}\right).
	\end{equation}
\end{theorem} 
\noindent According to the above theorem, there exists an estimator  $\hat{\beta}_{SL}$ of $\beta$ that is  root-$n/M_1M_2$ consistent.

Before stating Theorem \ref{the_3}, let us define with $\bm{b}_{ \left(1\right)}$ the vector whose entries are the $q$ non-zero elements of  $\bm{b}$ that are  and  with $\bm{b}_{\left(2\right)}$ the vector whose entries are the $(M_1+k_1)(M_2+k_2)-q$ elements of  $\bm{b}$ that are equal  to zero.
In what follows, we assume, without loss of generality, that $\bm{b}=\left[\bm{b}_{\left(1\right)}^{T}\quad \bm{b}_{\left(2\right)}^{T}\right]^{T}$ and that  a  matrix $\bm{A}_{l}\in \mathbb{R}^{(M_1+k_1)(M_2+k_2)\times (M_1+k_1)(M_2+k_2)}$ can be expressed in  block-wise form as 
\[
\bm{A}_{l}=\left[\begin{array}{c c}
\bm{A}_{l,11}\in \mathbb{R}^{q\times q}&\bm{A}_{l,12}\in \mathbb{R}^{q\times (M_1+k_1)(M_2+k_2)-q}\\
\bm{A}_{l,21}\in \mathbb{R}^{(M_1+k_1)(M_2+k_2)-q\times q}&\bm{A}_{l,22}\in \mathbb{R}^{(M_1+k_1)(M_2+k_2)-q\times (M_1+k_1)(M_2+k_2)-q}
\end{array}\right].
\]
To prove Theorem \ref{the_3}, in addition to C.\ref{as_0}-C.\ref{as_2p}, the following conditions are considered.
\begin{assumption}
	\label{as_5}
	(S-LASSO irrepresentable condition (SL-IC)) There exists $\lambda_s$, $\lambda_t$, $\lambda_{L}$, and a constant $\eta>0$ such  that, element-wise,
	\begin{align*}
	\Big|\bm{W}_{st,21}	^{-1}\Big\{\left[\left(\bm{W}_{t}\otimes n^{-1}\bm{X}^{T}\bm{X}\right)_{21}+n^{-1}\lambda_{s}\bm{L}_{wr,21}+n^{-1}\lambda_{t}\bm{L}_{rw,21}\right]\\
	\left[\left(\bm{W}_{t}\otimes n^{-1}\bm{X}^{T}\bm{X}\right)_{11}+n^{-1}\lambda_{s}\bm{L}_{wr,11}+n^{-1}\lambda_{t}\bm{L}_{rw,11} \right]^{-1}\\
	\left[\bm{W}_{st,11}	\sign\left(\bm{b}_{\alpha\left(1\right)}\right)+2\lambda_L^{-1}\lambda_{s}\bm{L}_{wr,11}\bm{b}_{\left(1\right)}+2\lambda_L^{-1}\lambda_{t}\bm{L}_{rw,11}\bm{b}_{\left(1\right)}\right]\\
	-2\lambda_L^{-1}\lambda_{s}\bm{L}_{wr,21}\bm{b}_{\left(1\right)}-2\lambda_L^{-1}\lambda_{t}\bm{L}_{rw,21}\bm{b}_{\left(1\right)}\Big\}\Big|\leq 1-\eta.
	\end{align*}
\end{assumption}
\begin{assumption}
	\label{as_6}
	The functions $\varepsilon_{i}\left(t\right)$ in Equation \eqref{eq_lm} are zero mean Gaussian random processes with autocovariance function $K\left(t_1,t_2\right)$, $t_1$ and $t_2 \in \mathcal{T}$, independent of $X_{i}$.
\end{assumption}
\begin{assumption}
	\label{as_7}
	Given $\rho\doteq\min|\left[\left(\bm{W}_{t}\otimes\bm{X}^{T}\bm{X}\right)_{11}\right.\allowbreak\left.+\lambda_{s}\bm{L}_{wr,11}+\lambda_{t}\bm{L}_{rw,11} \right]^{-1}\left[\left(\bm{W}_{t}\otimes\bm{X}^{T}\bm{X}\right)_{11}\bm{b}_{\left(1\right)}\right]|$ and $C_{min}\doteq\Lambda_{min}\left[\left(\bm{W}_{t}\otimes n^{-1}\bm{X}^{T}\bm{X}\right)_{11}\right]$,  $\Lambda_{min}\left(\bm{W}_{t}\right)M_2\rightarrow c_w$ as $n\rightarrow\infty$, with $0<c_w<\infty$, and  the parameters $\lambda_s$, $\lambda_t$ and $\lambda_{L}$ are chosen such that
	\begin{enumerate}
		\item
		\begin{equation*}
		\frac{M_{1}^{2}M_{2}^{2}\log\left[\left(M_1+k_1\right)\left(M_2+k_2\right)-q\right]}{\lambda_L^2}\left[nc^{2}+\frac{\lambda_{s}^{2}\Lambda_{max}^{2}\left(\bm{L}_{wr}\right)}{nC_{min}}+\frac{\lambda_{t}^{2}\Lambda_{max}^{2}\left(\bm{L}_{rw}\right)}{nC_{min}}\right]=o\left(1\right),
		\end{equation*}
		\item
		\begin{align*}
		&\frac{1}{\rho}\Big\{ \sqrt{\frac{M_1M_2\log\left(q\right)}{nC_{min}}}\\
		&+\frac{\lambda_L}{nM_1M_2}\Lambda_{min}^{-1}\left[\left(\bm{W}_{t}\otimes n^{-1}\bm{X}^{T}\bm{X}\right)_{11}+\lambda_{s}n^{-1}\bm{L}_{wr,11}+\lambda_{t}n^{-1}\bm{L}_{rw,11} \right]||\sign\left(\bm{b}_{\left(1\right)}\right)||_{2}\Big\}=o\left(1\right).  
		\end{align*}
	\end{enumerate}
\end{assumption}
\noindent The SL-IC in C.\ref{as_5} is the straightforward generalization to the problem in Equation \eqref{eq_slasso} of the elastic irrepresentable condition described in \cite{jia2010model}. It is a consequence of the  standard Karush$-$Kuhn$-$Tucker (KKT) conditions applied to the optimization problem in Equation \eqref{eq_slassoobj}. C.\ref{as_6} gives some conditions on the relationship of $\lambda_s$, $\lambda_t$, and  $\lambda_{L}$ with $M_1$, $M_2$ and $n$.
In the classical setting, an estimator is sign selection consistent if it has the same sign of the true parameter with probability tending to one. Analogously, we say that an estimator of the coefficient function $\beta$ is pointwise sign consistent if, in each point of the domain, it has the same sign of $\beta$ with probability tending to one.  The following theorem states that, under opportune assumptions, the S-LASSO estimator is pointwise sign consistent.
\begin{theorem}
	\label{the_3}	
	Under assumptions C.\ref{as_0}-C.\ref{as_2p} and  C.\ref{as_5}-C.\ref{as_7},  $\hat{\beta}_{SL}$ is pointwise sign consistent, that is, for all $s\in\mathcal{S}$ and $t\in\mathcal{T}$,
	
	\begin{equation}
	\Pr\Big\{\sign\left[\hat{\beta}_{SL}\left(s,t\right)\right]=\sign\left[\beta\left(s,t\right)\right]\Big\}\rightarrow 1,
	\end{equation}
	as $n\rightarrow\infty$.
\end{theorem} 
This theorem is the functional extension of the sign consistency result for the multivariate LASSO estimator \citep{zou2009adaptive}.
\begin{remark}
	\label{re_3}
	Assumption C\ref{as_0} is in fact not in contradiction with  the assumption that the null space of $ K_X $, i.e., the covariance operator   of $ X_i $,  is empty.
	While the assumption that the null space of $ K_X $ is related to the joint variability of $X$ over the domain $\mathcal{S}$, the assumption C\ref{as_0} refers to the magnitude of $X$, and, thus, broadly speaking, to its expected value. Following \cite{bosq2000linear}, $K_X$ can be decomposed as 
	\begin{equation*}
		K_X f(s')=\sum_{j=1}^{\infty}\lambda_j\left(\int_{\mathcal{S}}v_j(s)f(s)ds\right)v_j(s') \quad s'\in \mathcal{S}, 
	\end{equation*}
	where $\left( \lambda_j,v_j\right)$, $j\geq 1$ is a complete sequence of eigenelements of $K_X$ such that $\sum_{j=1}^{\infty}\lambda_j=E||X||_2^2<\infty$. From assumption C\ref{as_0} we know that $E||X||_2^2<c^2$ almost surely. Thus, assumption C\ref{as_0} provides information ust on the eigenvalues sum and not on their single  values. The null space of $ K_X $ is empty when all the eigenvalues are strictly larger than zero, while assumption C\ref{as_0} is not related to single  eigenvalues.
	For these reasons, these assumptions are simultaneously used in several papers \citep{cardot2003spline,prchal2007spline,zhou2012spline}.
	\end{remark}
\section{Simulation Study}
\label{sec_sim}
\begin{figure}

	\begin{subfigure}[b]{0.32\textwidth}
		
		\centering
		\includegraphics[width=\textwidth]{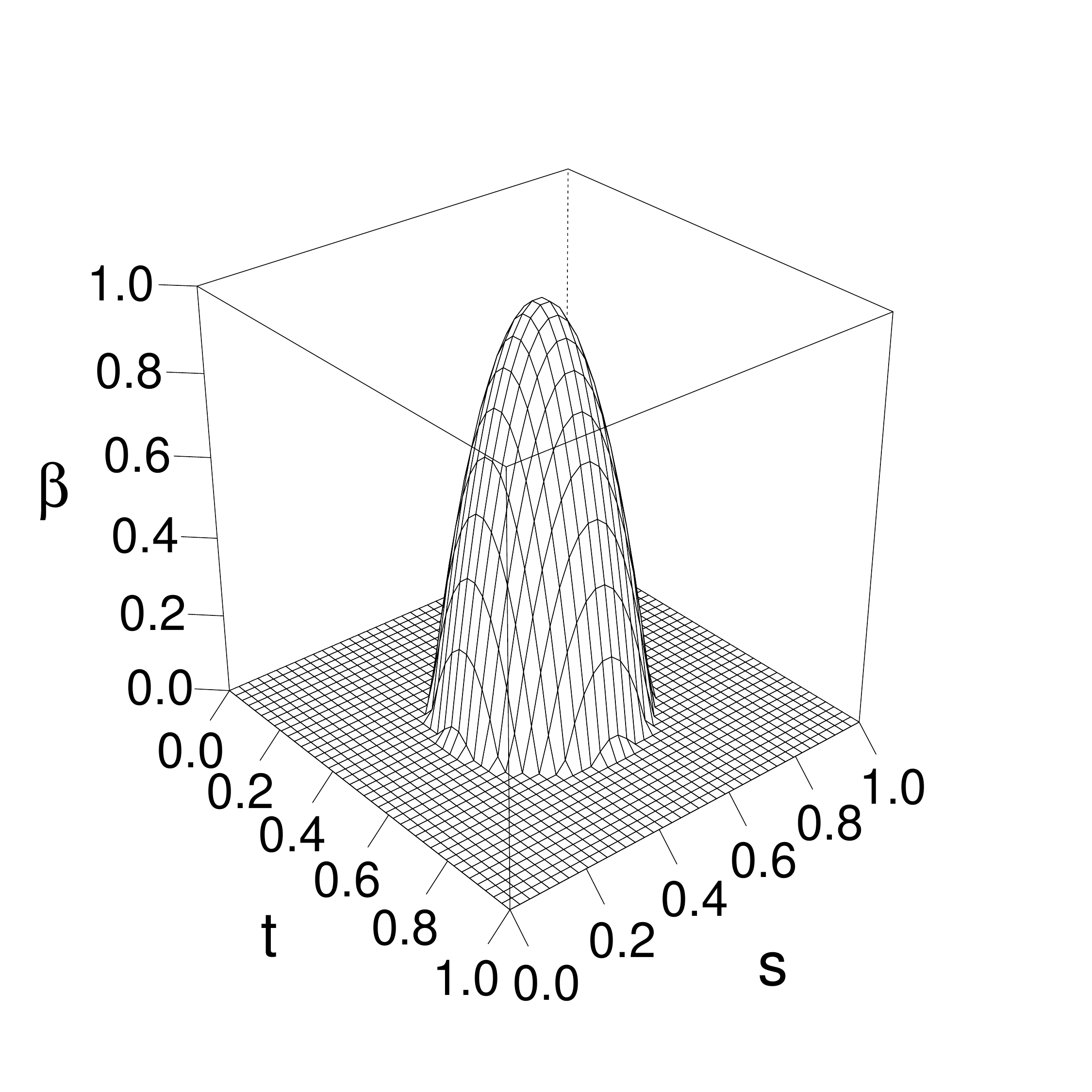}
		\caption{}
		\label{subfig_beta_2}
	\end{subfigure}
	\begin{subfigure}[b]{0.32\textwidth}
		\includegraphics[width=\textwidth]{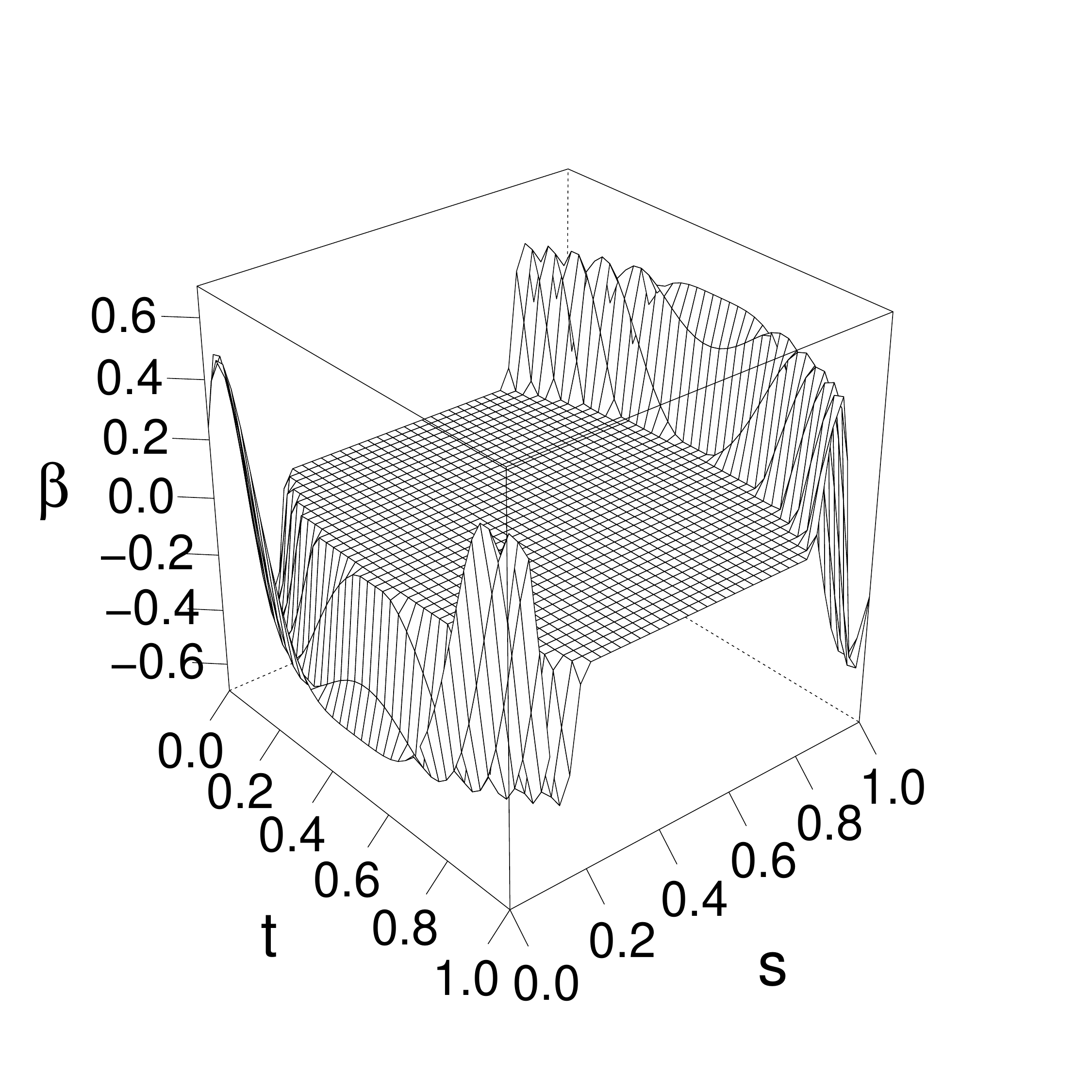}
		\caption{}
		\label{subfig_beta_3}
	\end{subfigure}
	\begin{subfigure}[b]{0.32\textwidth}
		\includegraphics[width=\textwidth]{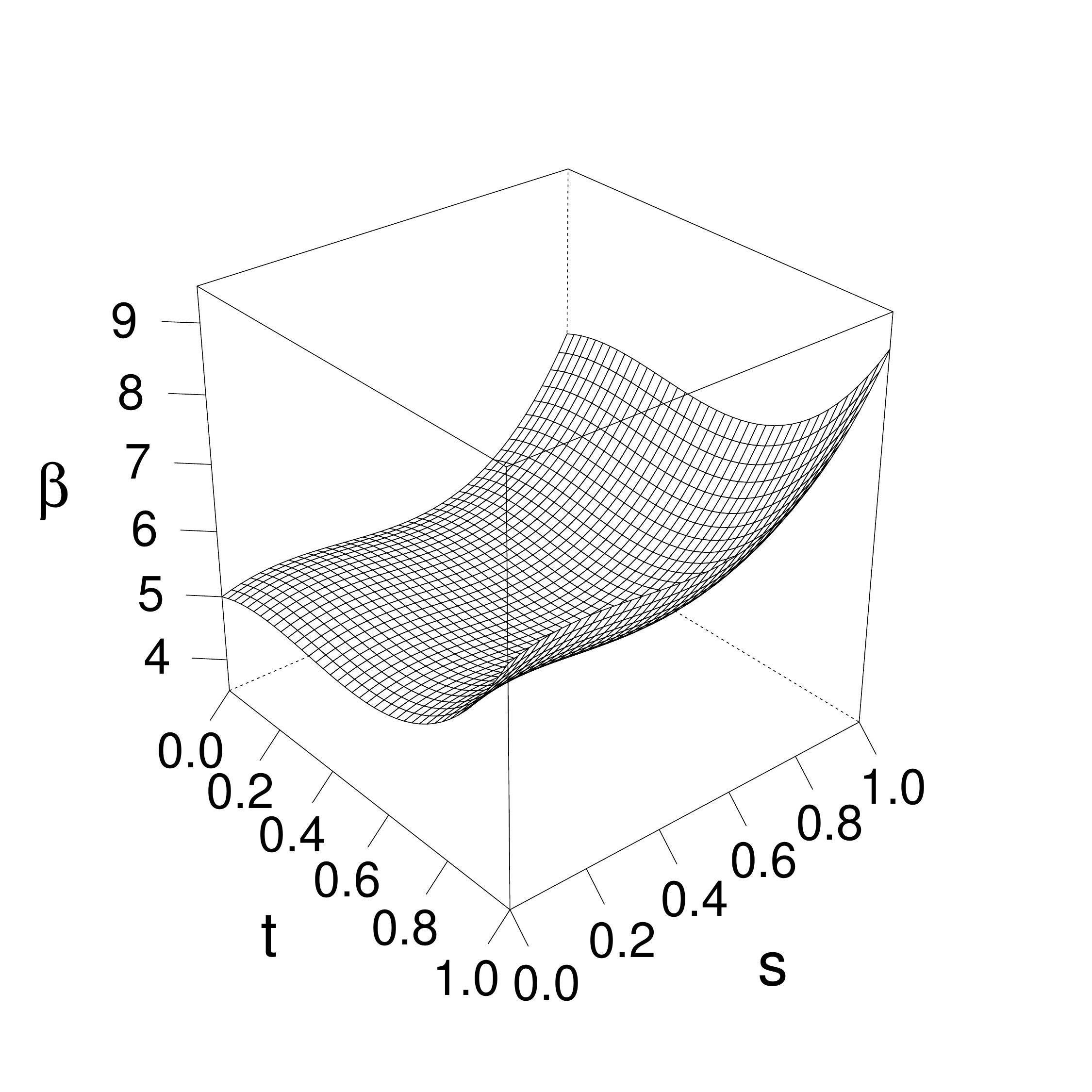}
		\caption{}
		\label{subfig_beta_4}
	\end{subfigure}
	\caption{True coefficient function $\beta$ for Scenario II \subref{subfig_beta_2}, Scenario III \subref{subfig_beta_3} and Scenario IV \subref{subfig_beta_4} in the simulation study. }
	\label{fig_betasim}
	
\end{figure}

In this section,  we conduct a Monte Carlo simulation study to explore the performance of the S-LASSO estimator.
We consider  four scenarios whose corresponding coefficient functions are depicted in Figure \ref{fig_betasim}.
Note that the coefficient function for Scenario I is not shown because it is zero all over the domain.
In Scenario II and III, $ \beta $ is sparse, indeed, it is zero on the edge and in the central part of the domain, respectively.
Scenario IV corresponds to a non-sparse setting, which is not expected to be favourable to the S-LASSO estimator.
The independent observations of the covariates  $X_{i}$  are generated as $X_{i}=\sum_{j=1}^{32}x_{ij}\psi_{i}^{x}$ where the coefficients $x_{ij}$  are independent realizations of truncated standard normal random variable defined on the interval $\left[-1000,1000\right]$, and $\psi_{i}^{x}\left(s\right)$ are cubic  B-splines with evenly spaced knot sequence.
Further details on the data generation  are given in the Supplementary Material C.

For each scenario, we generate 100 datasets composed of a training set with sample size $n$  and a test set $T$ with  size $N$ equal to 4000 that are used to estimate the coefficient function and to test its predictive performance. This is repeated for three different sample sizes  $n=150,500,1000$.  
As in \cite{lin2017locally}, we consider the integrated squared error (ISE) to asses the quality of the estimator $\hat{\beta}$ of the coefficient function $\beta$.
In particular, the ISE over the null region (ISE\textsubscript{0}) and   the non-null region (ISE\textsubscript{1}) are defined as
\begin{equation}
\text{ISE\textsubscript{0}}=\frac{1}{A_{0}}\int\int_{N\left(\beta\right)}\left(\hat{\beta}\left(s,t\right)-\beta\left(s,t\right)\right)^{2}dsdt ,
\quad
\text{ISE\textsubscript{1}}=\frac{1}{A_{1}}\int\int_{NN\left(\beta\right)}\left(\hat{\beta}\left(s,t\right)-\beta\left(s,t\right)\right)^{2}dsdt,
\end{equation}
where $A_{0}$ and $A_{1}$ are the measures of the null ($N\left(\beta\right)$) and non-null ($NN\left(\beta\right)$) regions, respectively.
The ISE\textsubscript{0} and the ISE\textsubscript{1} are indicators of the estimation error of $\hat{\beta}$  over both the null and the non-null regions.
Moreover, predictive performance is measured through the prediction mean squared error (PMSE), defined as
\begin{equation}
	\label{eq_ise}
\text{PMSE}=\frac{1}{N}\sum_{\left(X,Y\right)\in T} \int_{0}^{1}\left(Y\left(t\right)-\int_{0}^{1}X\left(s\right)\hat{\beta}\left(s,t\right)ds\right)^{2}dt,
\end{equation}
where $\hat{\beta}$ is obtained through the observations in the training set. The observations in the test set are centred by means of  the sample mean functions estimated through  the training set observations.

As a remark, the coefficient function $ \beta $ is not identifiable in  $L^2 (\mathcal{S}\times\mathcal{T})$, because the $ X_i $ is generated as a finite linear combination  of basis functions, i.e., not empty  null space  of $ K_X $. Thus,  as stated in Section \ref{sec_theo}, $ \beta  $ is identifiable in the closure of $ Im(K_x) $.
To obtain a meaningful measure of the estimation accuracy, both $ ISE_0 $ and $ ISE_1 $ should be computed by considering estimate projections onto $ Im(K_x) $. This means,  the methods are compared by considering their estimation performance over the identifiable part of the model, only.
However, according to the works of \cite{james2009functional,zhou2013functional,lin2017locally},   the space spanned by the 32 cubic B-splines used to generate $ X_i $ is sufficiently rich to well approximate both $ \beta $ and $ \hat{\beta} $ for the proposed and competing methods. Therefore,  $ ISE_0 $ and $ ISE_1 $ in Equation \eqref{eq_ise} can be suitably used to assess the estimation  error of the coefficient function over $N\left(\beta\right)$ and $NN\left(\beta\right)$, respectively.

The S-LASSO estimator  is compared with four different estimators of $\beta$ which are already present in the literature of the FoF linear regression model estimation.
The first two are those proposed by \cite{ramsay2005functional}, where the coefficient function  estimator  is assumed to be in a finite dimension tensor space with regularization achieved either  by choosing the dimension of the tensor space or by introducing roughness penalties.
They will be referred to as TRU and SMOOTH estimators, respectively. 
The third one is the estimator proposed by \cite{ivanescu2015penalized,scheipl2015functional} (referred to as PFFR), which is implemented in the \textsf{pffr} function of the \textsf{R} package \textsf{refund}.
The fourth and fifth ones are those proposed by \cite{yao2005regression}, based on the functional principal components analysis (referred to as PCA), and by \cite{canale2016constrained}, based on a ridge-type penalization (referred to as RIDGE).
The TRU, SMOOTH and S-LASSO are computed by using cubic B-splines with evenly space knot sequences. The dimensions of the B-spline sets that generate the tensor product space for the SMOOTH and S-LASSO estimator are both set equal to 60.
Additional results  for different choices of the dimensions of the B-spline sets, as well as computational times,  are provided in  the Supplementary Material D.
The tuning parameters of the TRU, SMOOTH, PCA, and RIDGE  estimators are  chosen by means of  $10$-fold cross-validation, viz., the  dimension of the tensor basis space for the TRU, the roughness penalties for the SMOOTH, the numbers of retained principal components for the PCA, the penalization parameter for the RIDGE  and $\lambda_s$, $\lambda_t$ and $\lambda_{L}$ for the S-LASSO. In particular the $10$-fold cross-validation for the S-LASSO method is applied with the $0.5$-standard deviation rule. Finally, the PFFR estimator is computed through tensor
products of cubic B-splines with 15 basis functions and second-order
difference penalties in both directions with smoothing parameters estimated using restricted maximum likelihood (REML).

\begin{figure}
	
	\centering
	\includegraphics[width=0.43\textwidth]{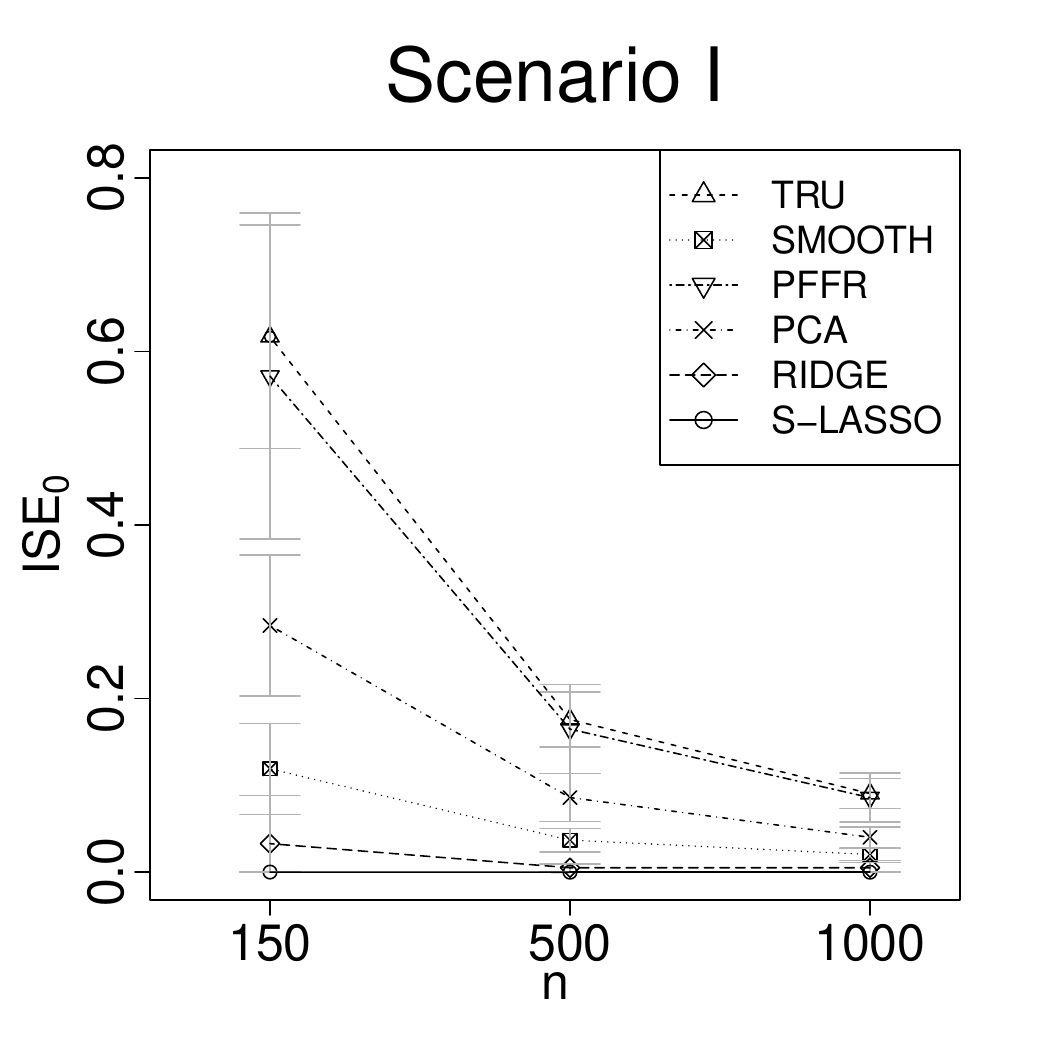}
	\includegraphics[width=0.43\textwidth]{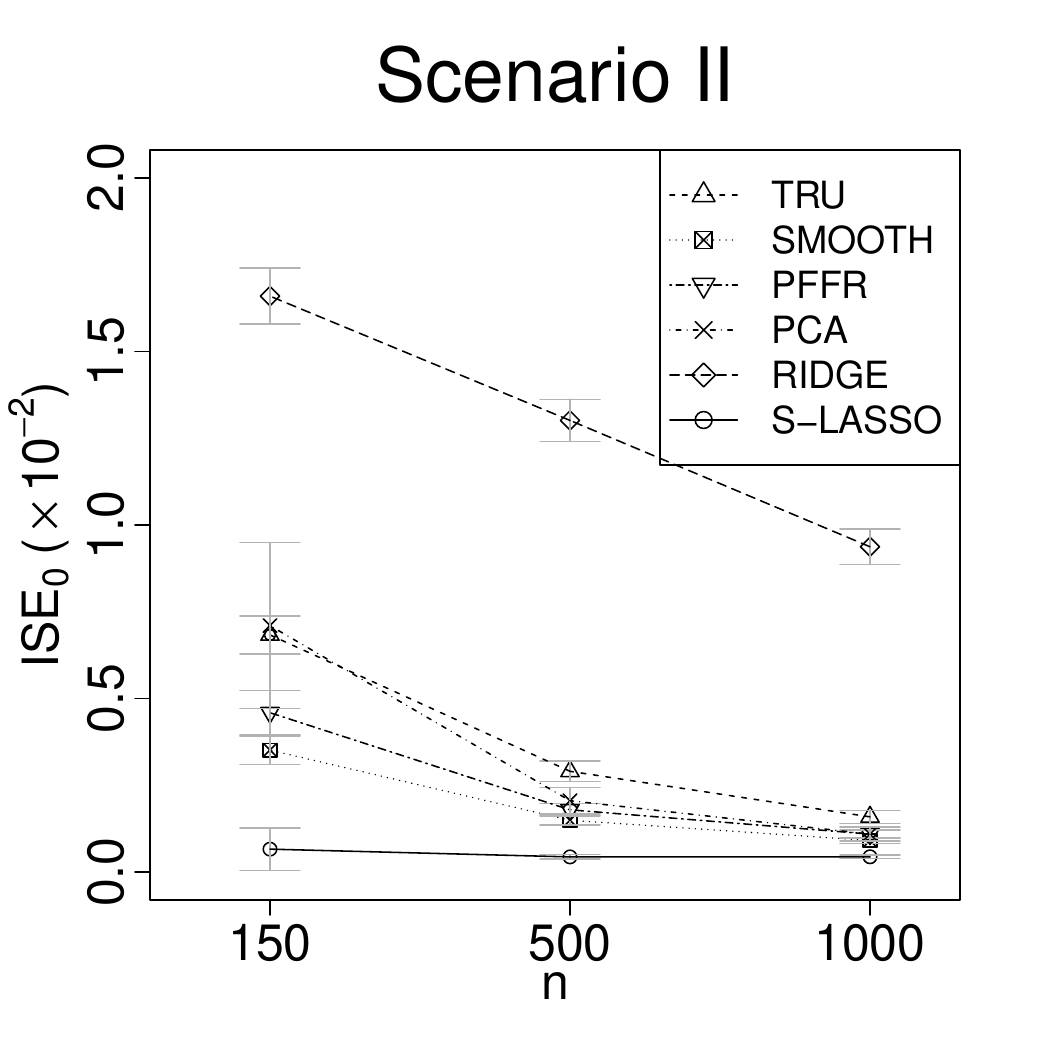}
	\includegraphics[width=0.43\textwidth]{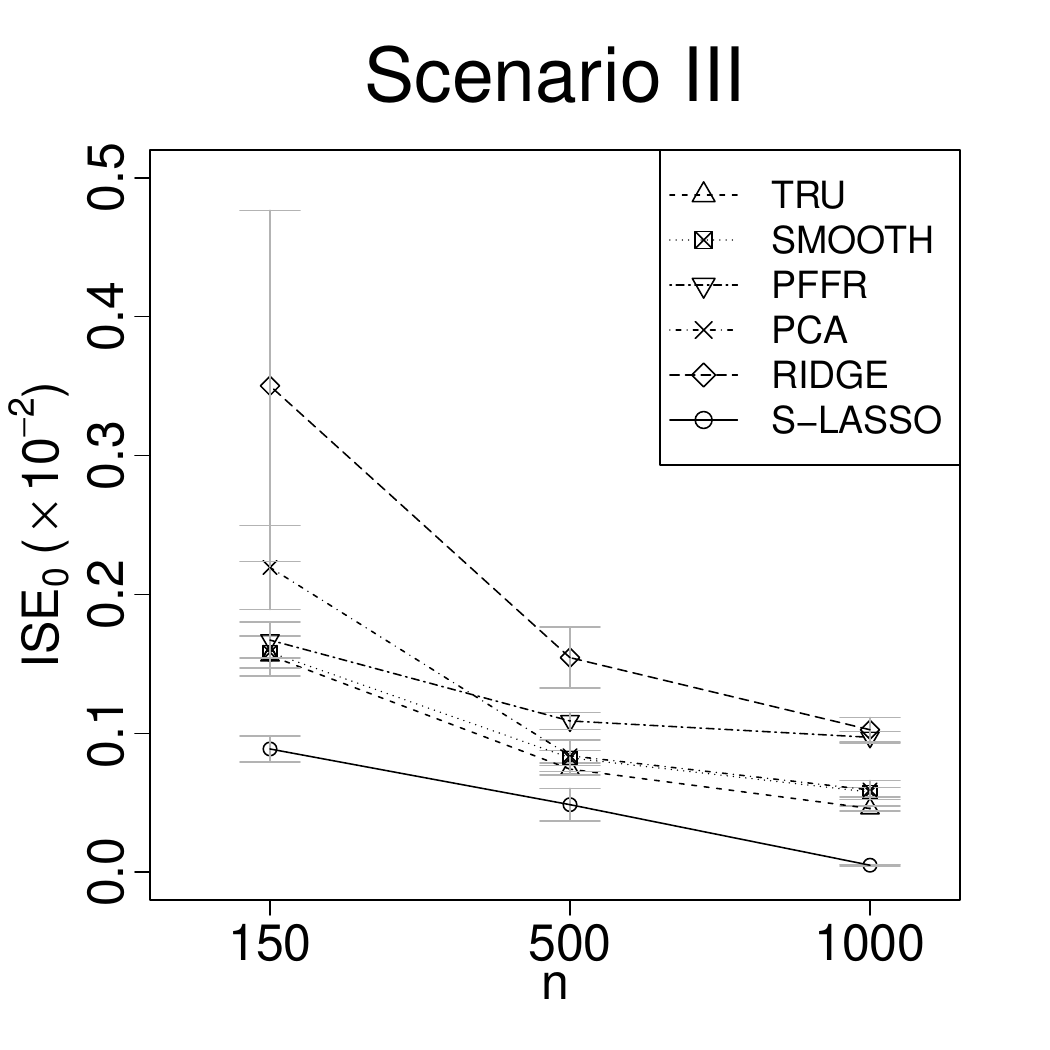}
	\caption{ The integrated squared error on the null region (ISE\textsubscript{0}) along with $\pm 0.5(standard\hspace{0.12cm} error)$   for the TRU, SMOOTH, PFFR, PCA, RIDGE, and S-LASSO  estimators.}
	\label{fig_ISE0}
	
\end{figure}

The performance of the estimators in terms of ISE\textsubscript{0} is displayed in Figure \ref{fig_ISE0}.
It is not surprising that the estimation error of $\beta$ over  $N\left(\beta\right)$ of the S-LASSO estimator is significantly smaller  than those of the other estimators, being the capability of recovering sparseness of $\beta$ its main feature.
In Scenario I, the RIDGE estimator is the only one that performs  comparably to   the S-LASSO estimator. This is in accordance with the multivariate setting where it is well known that, when the response is independent of the covariates, the ridge estimator is able to shrink all the coefficients towards zero. The TRU, SMOOTH, PFFR, and PCA estimators have difficulties to correctly identify $N\left(\beta\right)$ for all sample sizes. Nevertheless,  their performance is very  poor at $n=150$.
In Scenario II, the S-LASSO estimator is still the best one to estimate $\beta$ over  $N\left(\beta\right)$. However, in this case, the RIDGE estimator performance is unsatisfactory and is mainly caused by the lack of smoothness control that makes the estimator over-rough, especially at small $n$. Among the competitor estimators, the SMOOTH one has the best performance, readily followed by the PFFR estimator.
In Scenario III, results are similar to those of Scenario II, even if the TRU estimator appears as the best alternative. Both PCA and RIDGE estimators are  not able to successfully recover sparseness of $\beta$ for $n=150$. For the former, the cause is  the number of observations  not sufficient to capture the covariance structure of the data, whereas for the latter, it is due to the excessive roughness of the estimator.  For $n=500$ and $ m=1000 $, the PFFR estimator performs very poorly. This is due to the incapacity of the REML approach to appropriately select the smoothing parameters.

\begin{figure}
	
	\centering
	\includegraphics[width=0.43\textwidth]{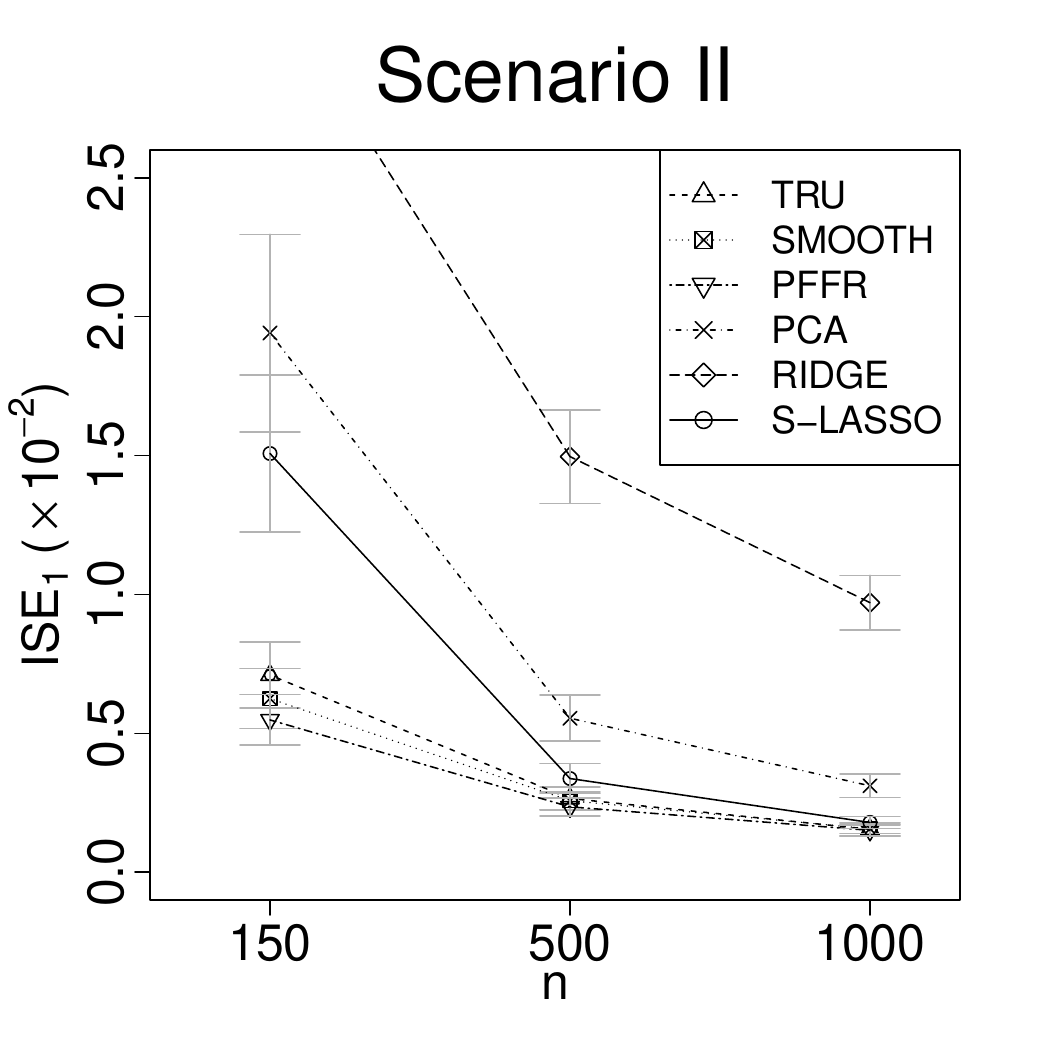}
	\includegraphics[width=0.43\textwidth]{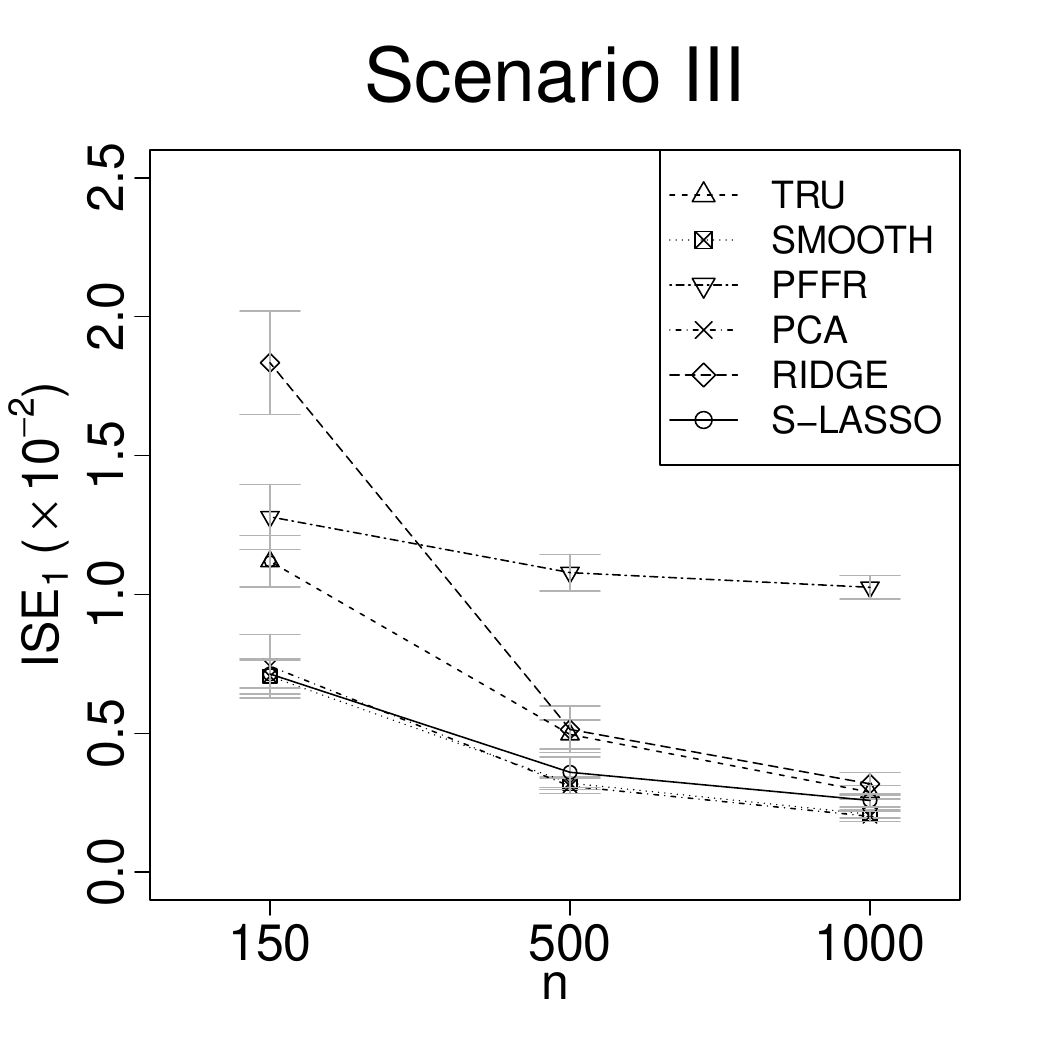}
	\includegraphics[width=0.43\textwidth]{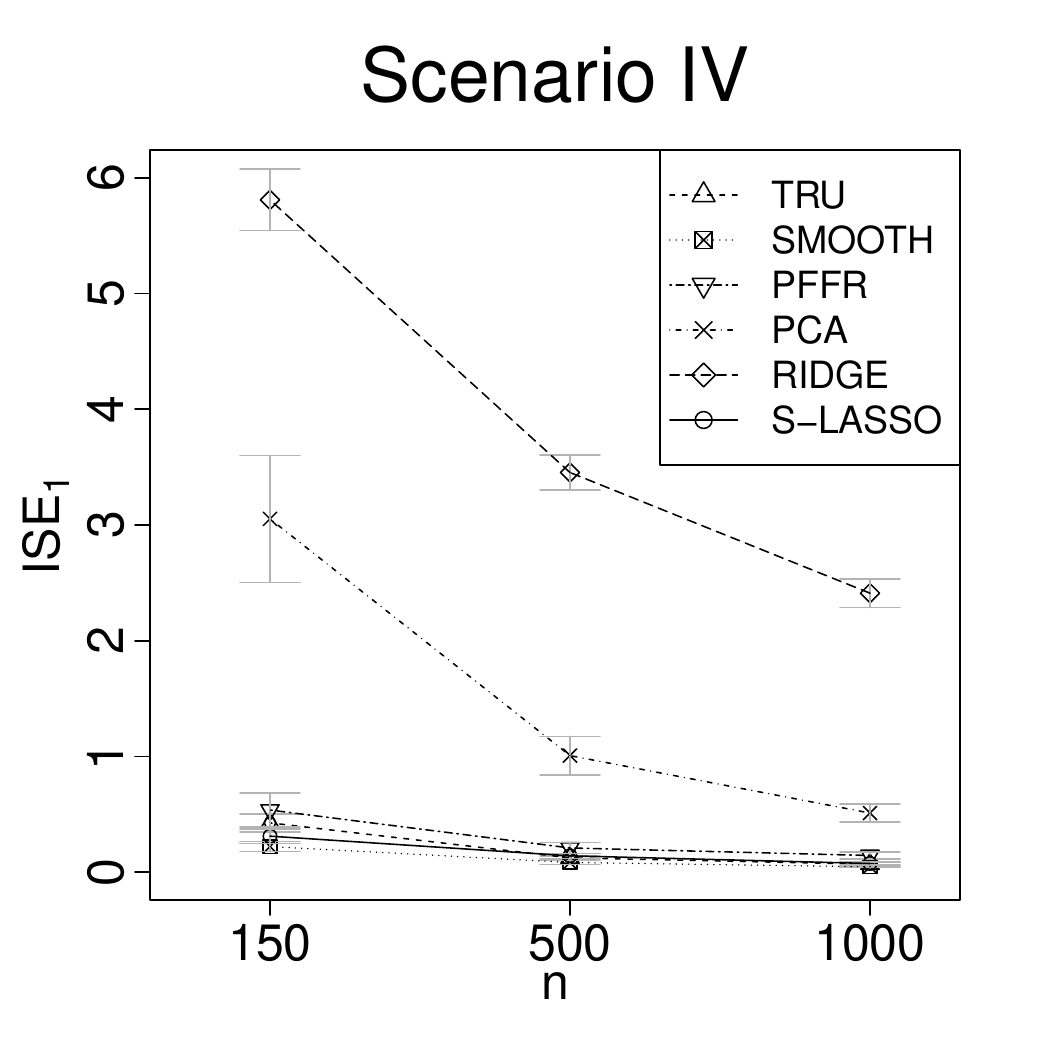}
	\caption{The integrated squared error on the non-null region (ISE\textsubscript{1}) along with $\pm 0.5(standard\hspace{0.12cm} error)$   for the TRU, SMOOTH, PFFR, PCA, RIDGE,
		 and S-LASSO  estimators.}
	\label{fig_ISE1}
	
\end{figure}

Results in terms of ISE\textsubscript{1} are summarized in Figure \ref{fig_ISE1}.
It is worth noting that, in this case, as expected the performance of the S-LASSO estimator is generally worse than that of the SMOOTH estimator. In some cases, it is worse than that of the TRU  and PFFR estimators as well.
However, in Scenario II performance differences between the S-LASSO estimator and  TRU, SMOOTH or PFFR estimators become negligible as sample size increases. The PCA and RIDGE estimators are always less efficient.
The results are similar for Scenario III, where  the performance of the S-LASSO estimator is  comparable  with that of the SMOOTH estimator. Except for the PFFR estimators that badly  performs due the difficulties of the REML approach to select the appropriate smoothing parameters, by comparing to the classical LASSO method, the behaviour of the S-LASSO estimator --- in terms of ISE\textsubscript{1} --- is not surprising. Indeed, it is well known that LASSO method does nice variable selection, even if it tends to overshrink the estimators of the non-null coefficients \citep{fan2004nonconcave,james2009generalized}. By looking at the result for Scenario II and III, we surmise that this phenomenon arises in the FoF linear regression model as well.
Finally, in Scenario IV,  where $\beta$ is always different from zero, the S-LASSO estimator,performs comparably to the SMOOTH (i.e., the S-LASSO estimator with $\lambda_{L}=0$). In this case $\beta$ is not sparse and, thus, the FLP does not help.

\begin{figure}
	
	\centering
	\includegraphics[width=0.43\textwidth]{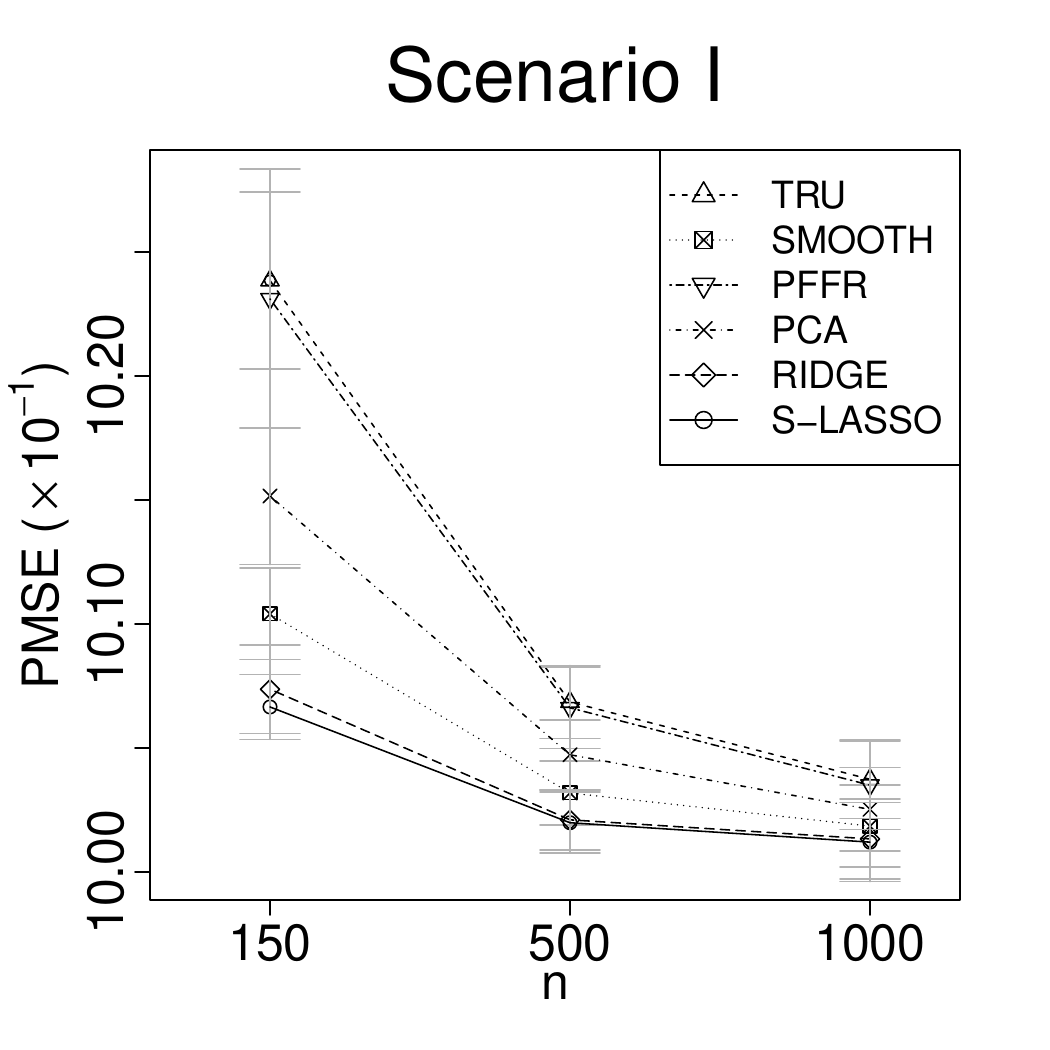}
	\includegraphics[width=0.43\textwidth]{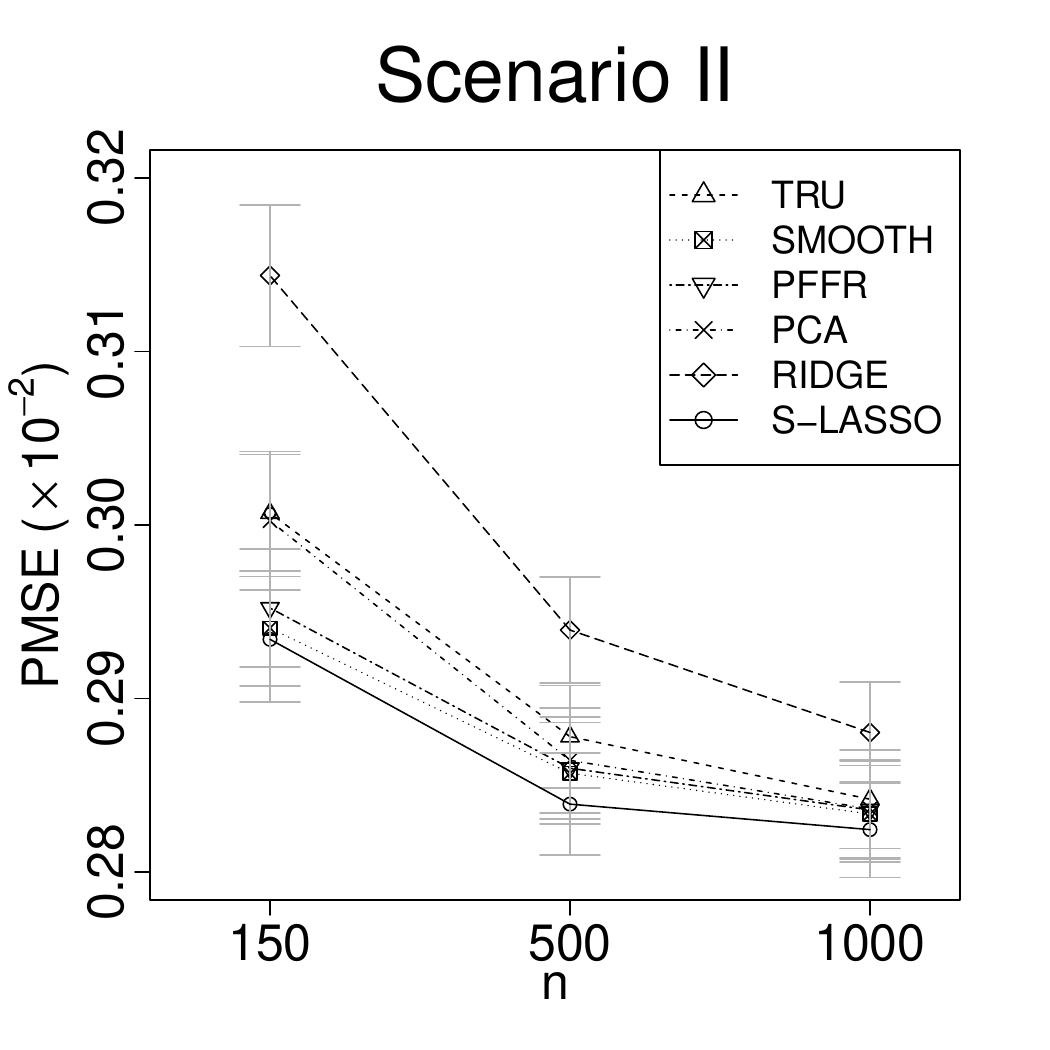}
	
	\includegraphics[width=0.43\textwidth]{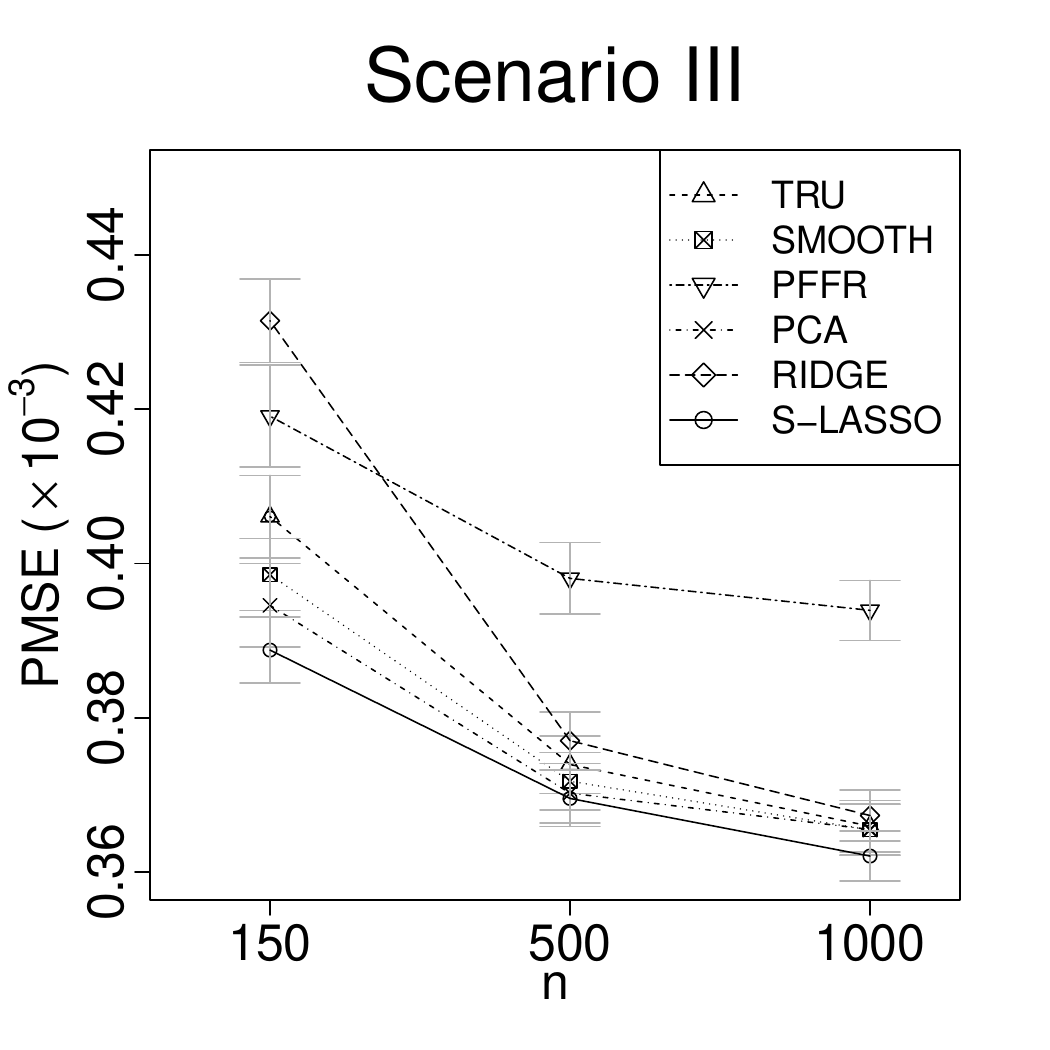}
	\includegraphics[width=0.43\textwidth]{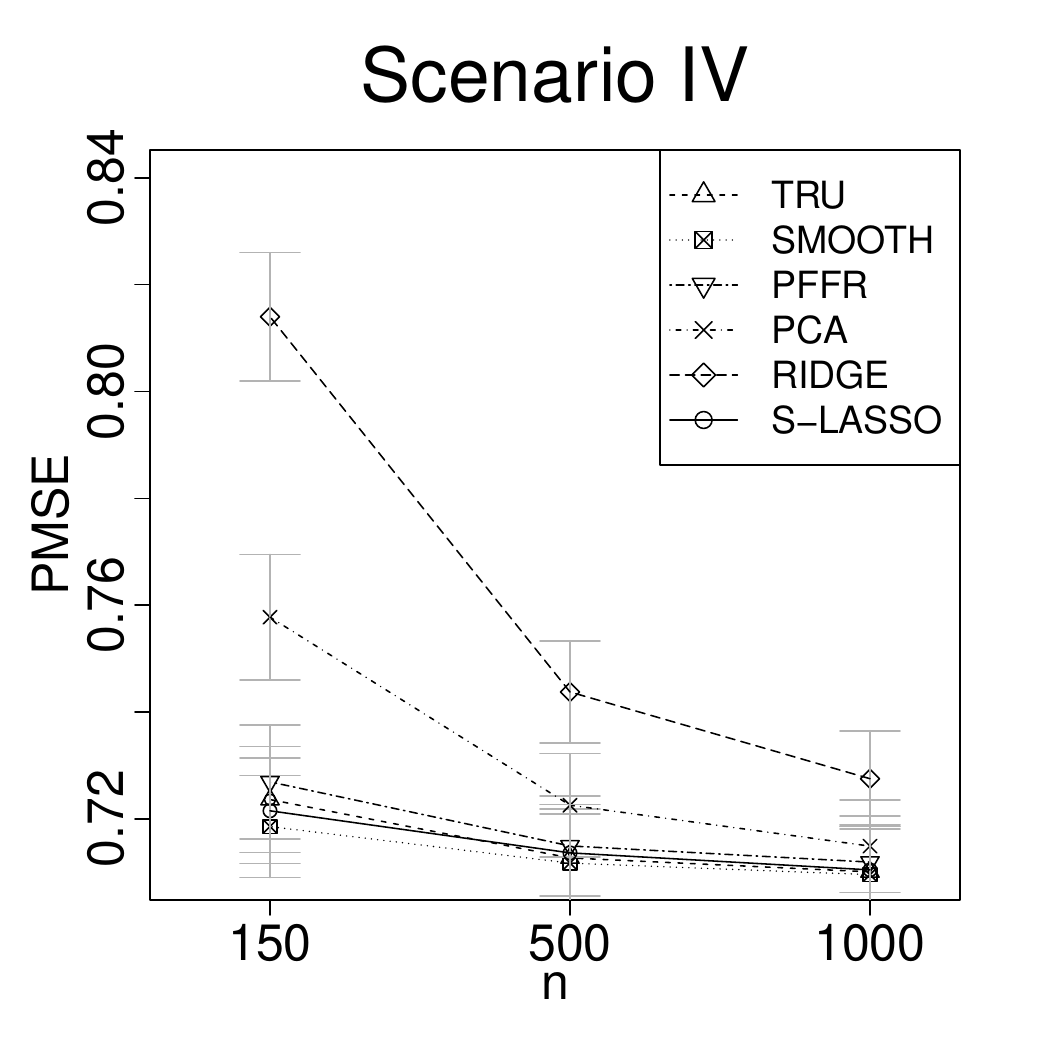}
	\caption{ The prediction mean squared error (PMSE) along with $\pm 0.5(standard\hspace{0.12cm} error)$   for the TRU, SMOOTH, PFFR, PCA, RIDGE, and S-LASSO estimators.}
	\label{fig_PMSE}
\end{figure}

Figure \ref{fig_PMSE} shows PMSE averages and corresponding standard errors for all the considered estimators.
Since PMSE is strictly related to the  ISE\textsubscript{0} and the ISE\textsubscript{1}, results are totally consistent with those of  Figure \ref{fig_ISE0} and  Figure \ref{fig_ISE1}. In particular, the S-LASSO estimator outperforms all the competitor ones in favorable scenarios (viz., Scenario I, II, and III),  being the corresponding PMSE lower than that achieved by the other competing estimators. 
In these scenarios, although the performance of the S-LASSO estimator in terms of ISE\textsubscript{1} is not excellent, the clear superiority in terms of  ISE\textsubscript{0} compensates and gives rise to smaller PMSE.
Otherwise, for Scenario IV, where the coefficient function is not sparse,  the performance of the S-LASSO estimator is very similar to that of the SMOOTH estimator, which is the best one in this case. This is encouraging, because, it proves that the performance of the  S-LASSO estimator does not dramatically decline in less favourable scenarios.

In summary, the S-LASSO estimator  outperforms the competitors  both in terms of estimation error on the null region and prediction accuracy on a new dataset, as well as that it is able to estimate competitively the coefficient function on the non-null region.
On the other hand, in order to achieve sparseness,	 the S-LASSO   tends to overshrink the estimator of the coefficient function on the non-null region.
This means that, as in the classical setting \citep{james2009generalized}, there is a trade-off between the ability of recovering sparseness and the estimation accuracy  on the non-null region of the final estimator. 
Moreover, even when the coefficient function is not sparse (Scenario IV), the proposed estimator demonstrates to have both  good prediction and estimation performance. This is another key property of the proposed estimator that,  encourages practitioners to use the S-LASSO estimator even when there is not prior knowledge about the shape of the coefficient function. Finally, it should be noticed that, in  scenarios similar to those analysed, the PCA and RIDGE estimators should not be preferred with respect to the TRU, SMOOTH and S-LASSO ones, and that the PFFR performance is strictly related to the REML approach to select the smoothing parameters.

\label{sec_sim}

\section{Real-Data Examples}
\label{sec_real}
In this section, we analyse two real-data examples.
We aim to confirm that the S-LASSO estimator has advantages in terms of both prediction accuracy and interpretability, over the SMOOTH estimator, which has been demonstrated in Section \ref{sec_sim} to be the best alternative among the competitors. The datasets used in the examples are the \textit{Canadian weather} and \textit{Swedish mortality}. Both are classical benchmark functional data sets thoroughly studied in the literature.
\subsection{Canadian Weather Data}
\label{sec_can}
\begin{figure}
	
	\centering
	\includegraphics[width=0.45\textwidth]{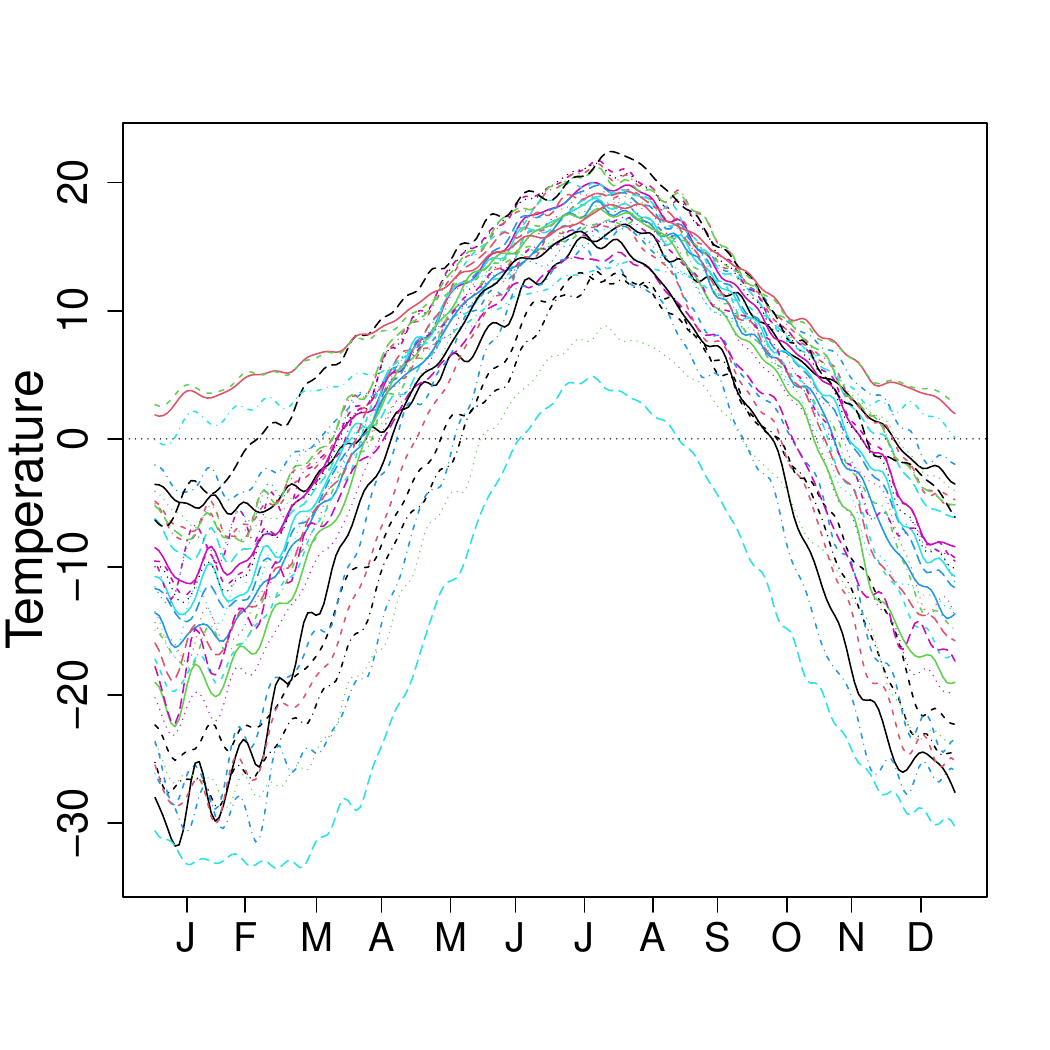}
	\includegraphics[width=0.45\textwidth]{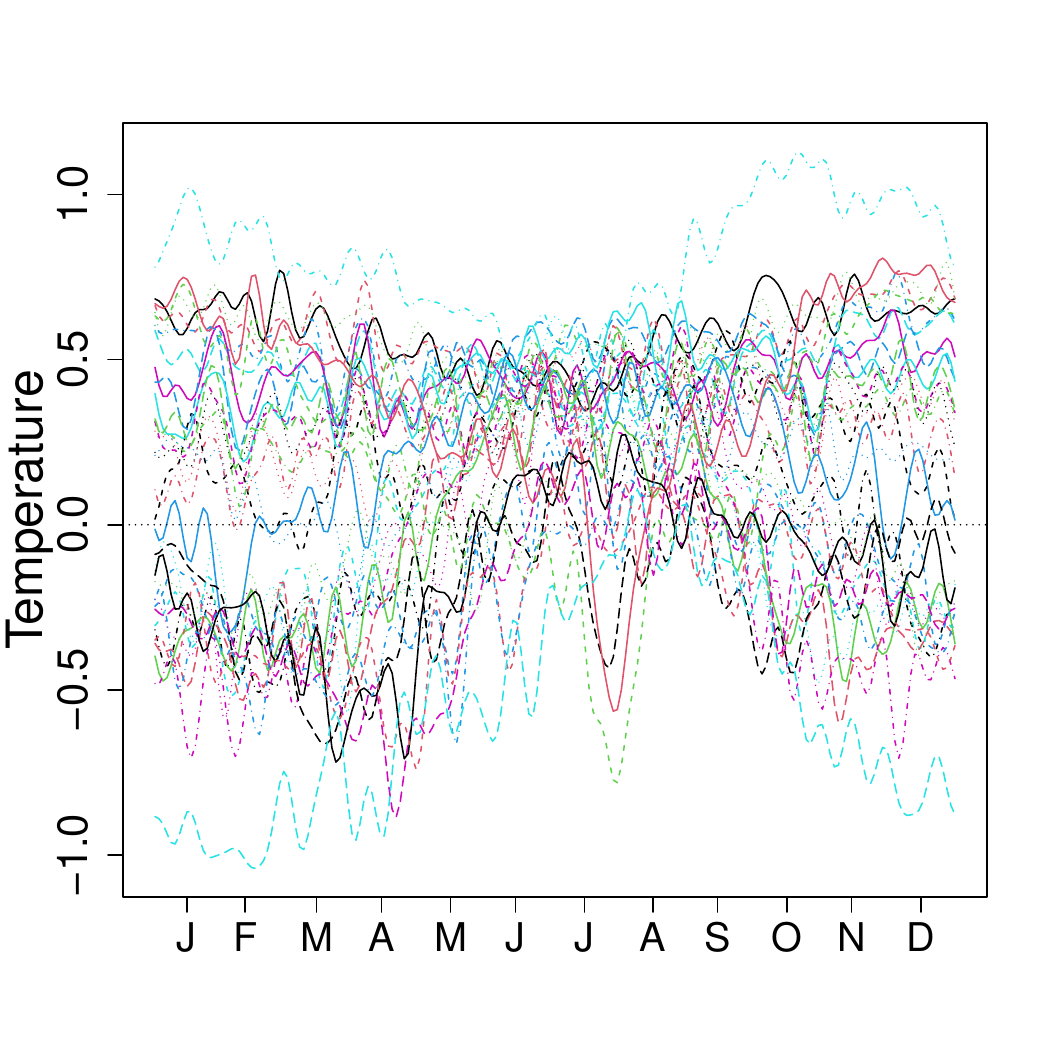}
	\caption{ Daily mean temperature and log-daily rainfall profiles at 35 cities in Canada over the year.}
	\label{fig_candata}
	
\end{figure}

\begin{figure}
	
	\centering
	\includegraphics[width=1\textwidth]{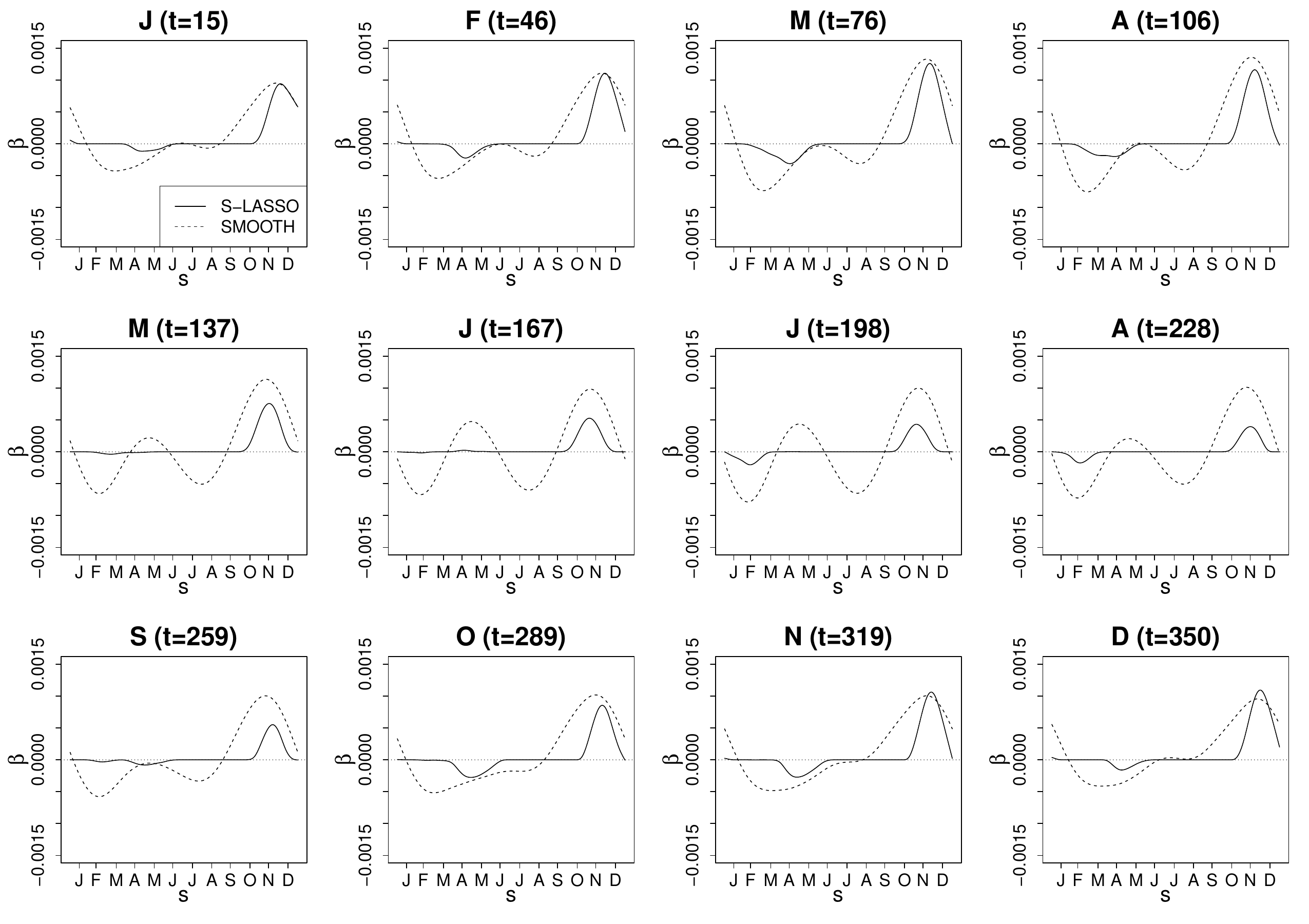}
	
	\caption{S-LASSO (\full) and SMOOTH (\dashed) estimates of the coefficient functions $\beta$  at different  months  for the Canadian weather data.}
	\label{fig_betacan}
	
\end{figure}

The Canadian weather data have been studied by \cite{ramsay2005functional} and \cite{sun2018optimal}.
The data set contains the  daily mean temperature curves, measured in Celsius degree,  and the log-scale of the  daily rainfall profiles, measured in millimeter,    recorded at 35 cities in Canada. Both temperature  and  rainfall profiles are obtained by averaging over  the years 1960 through 1994. Figure \ref{fig_candata} shows the profiles. 
The aim is to predict the log-daily rainfall based on the daily temperature using the model reported in  Equation \eqref{eq_lm}.
Figure \ref{fig_betacan} shows the S-LASSO and SMOOTH estimates of the coefficient function $\beta$.
The SMOOTH estimate is obtained using a Fourier basis---to take into account the periodicity of the data---and  roughness penalties were chosen by using 10-fold cross-validation over an opportune grid of values.
10-fold cross-validation is used to set the parameters $\lambda_s$, $\lambda_t$ and $\lambda_L$ as well.

The S-LASSO estimates is roughly zero over large domain portions. In particular, except for values from July through August, it is always zero in summer months (i.e., late June, July, August and September) and  in January and February. This suggests   in those months rainfalls  are not significantly influenced by daily temperature throughout the year. Otherwise, temperature in  fall months (i.e., October, November and December) gives strong positive contribution on the daily rainfalls. In other words, the higher (the lower) the temperature in October, November and December, the heavier  (the lighter) the precipitations throughout the year.
It is interesting  that the S-LASSO estimate in  spring months (i.e., March, April and May) is negative  for values of $t$ form January through April, and from October through December. This suggests that the higher (the lower) the temperature in the spring the lighter (the heavier) the daily rainfalls  from October through April. Finally, it is evidenced  a small influence of the temperature in February on  precipitations in July and August. It is worth noting that the S-LASSO estimate is more interpretable than the SMOOTH estimates, which does not allow for a straightforward interpretation.
Moreover, the S-LASSO estimate appears to have, even if slightly,  better prediction performance than the SMOOTH one. Indeed, 10-fold cross-validation mean squared errors are 22.314 and 22.365, respectively. 

\begin{figure}
	
	\centering
	\begin{subfigure}[b]{0.45\textwidth}
		
		\centering
		\includegraphics[width=\textwidth]{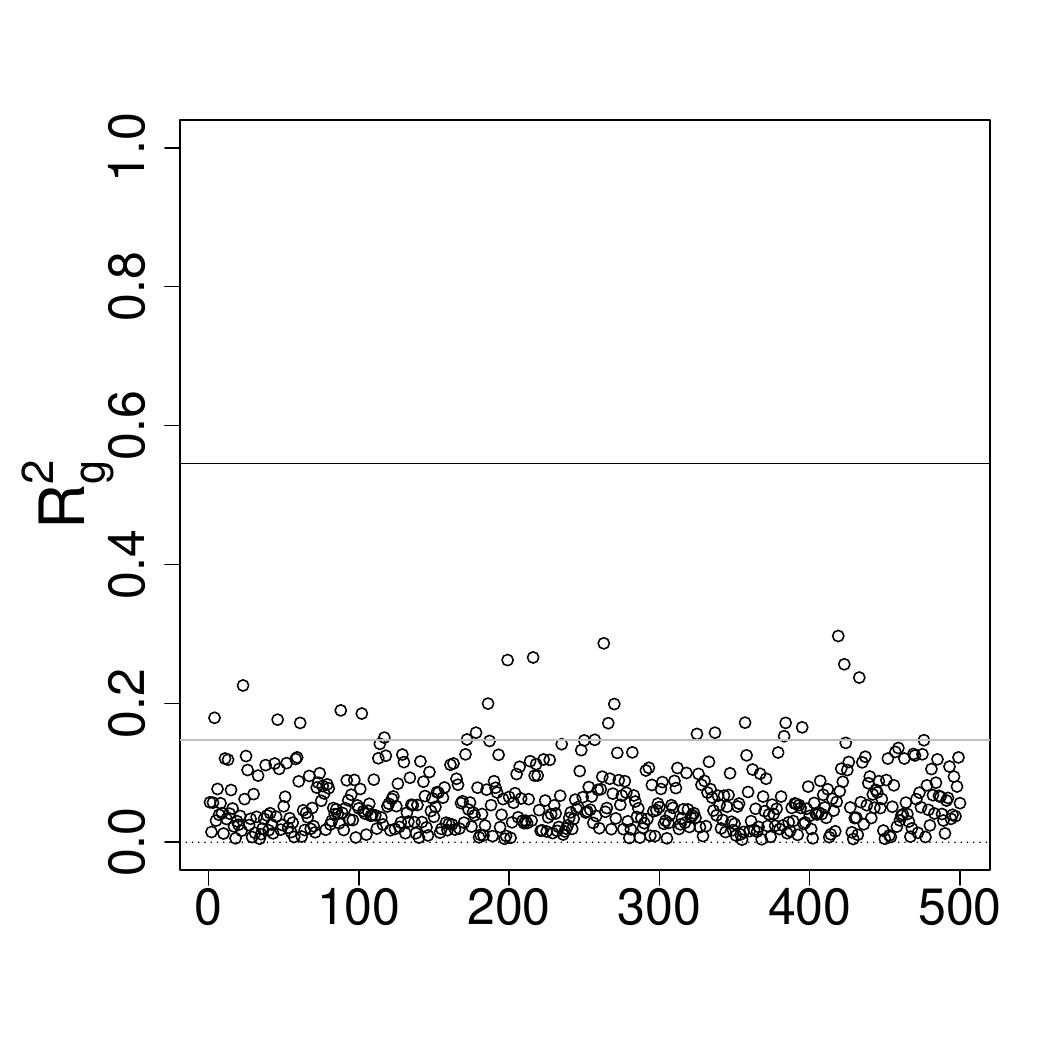}
		\vspace{-0.8cm}
		\caption{}
		\label{subfig_percan_1}
	\end{subfigure}
	\begin{subfigure}[b]{0.45\textwidth}
		
		\centering
		\includegraphics[width=\textwidth]{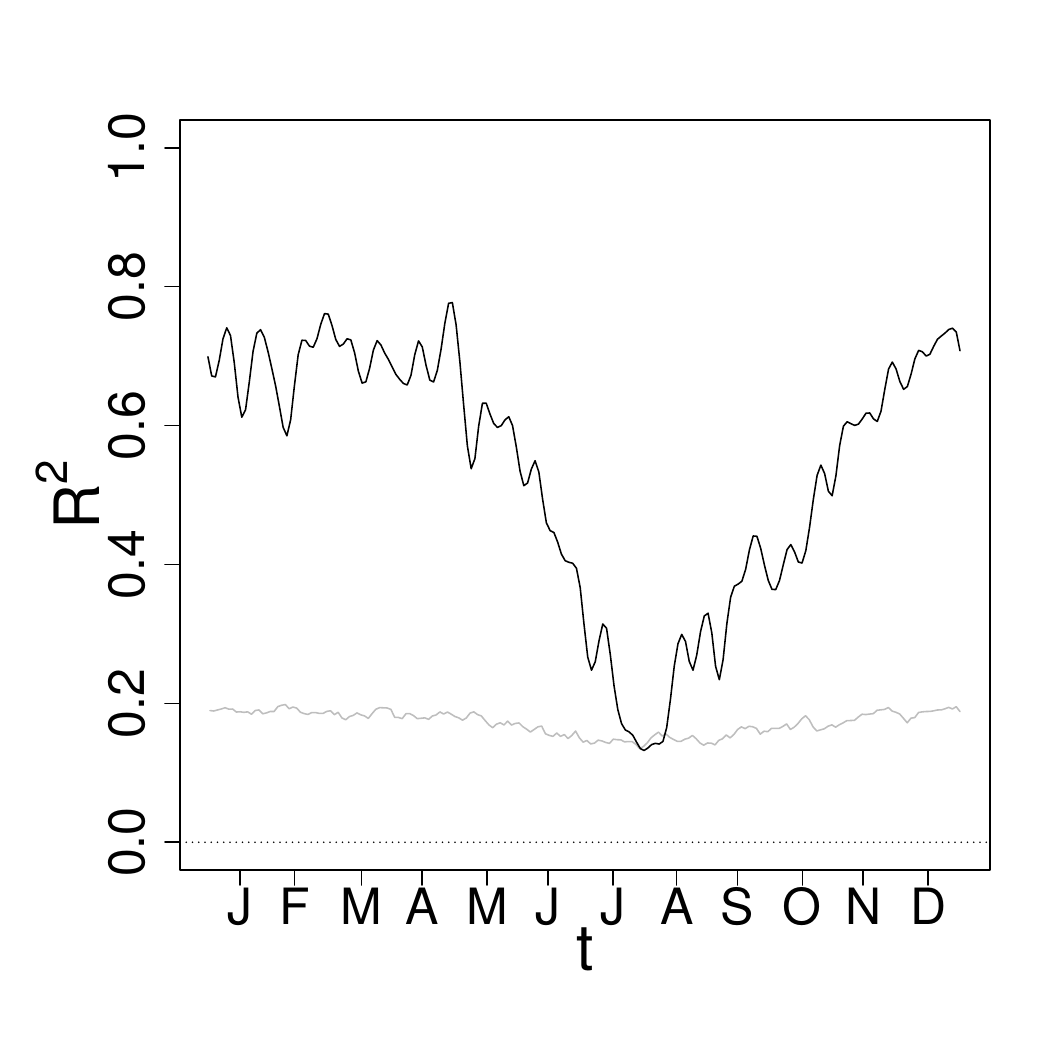}
		\vspace{-0.8cm}
		\caption{}
		\label{subfig_percan_2}
	\end{subfigure}

	\caption{For the Canadian weather data, \subref{subfig_percan_1} $R^{2}_{g}$ from permuting the response 500 times, where the black line corresponds to the observed $R^{2}_{g}$ and the grey line  to  the  $95$th sample percentile; \subref{subfig_percan_2} the black line is the observed $R^{2}$ and the grey line is the pointwise $95$th sample percentile curve. }
	\label{fig_percan}
	
\end{figure}

Finally, we perform two permutation tests to asses the statistical significance of the S-LASSO estimator.
The first test is based on the \textit{global functional coefficient of determination} defined as $R^{2}_{g}\doteq\int_{\mathcal{T}}\frac{\Var\left[\Ex\left(Y\left(t\right)|X\right)\right]}{\Var\left[Y\left(t\right)\right]}dt$ \citep{horvath2012inference}, with $\mathcal{T}=\left[0,365\right]$.  In Figure \ref{subfig_percan_1} the solid black line indicates the observed $R^{2}_{g}$ that is equal to 0.55. The bold points represent   500 $R^2_{g}$ values obtained by means of random permutations of the response variable. Whereas, the  grey line correspond to the  $95$th sample percentile.
All 100 values of $R^{2}$ as well as the value of the  $95$th sample percentile is far below 0.55, which gives a strong evidence of a significant relationship between  rainfalls and  temperature, globally.\\ 
By a second test, we aim to analyse the pointwise statistical significance, i.e., for each $t\in\mathcal{T}$. It is based on the \textit{pointwise functional coefficient of determination} defined as $R^{2}\left(t\right)\doteq\frac{\Var\left[\Ex\left(Y\left(t\right)|X\right)\right]}{\Var\left[Y\left(t\right)\right]}$ for $t\in\mathcal{T}$ \citep{horvath2012inference}.
Figure \ref{subfig_percan_2} shows the observed $R^2$ (solid black line) along with the pointwise $95$th sample percentile curve. The latter has been obtained by means of  500   $R^2$ values produced by randomly permuting the response variable. The observed $R^{2}$ is far above the 
$95$th sample percentile curve, except for some summer months  (viz., July and August). 
As global conclusion, we can state that the temperature has a large influence on the rainfalls in  autumn,  winter and spring.

\subsection{Swedish Mortality Data}
\label{sec_sw}
\begin{figure}
	
	\centering
	\includegraphics[width=0.45\textwidth]{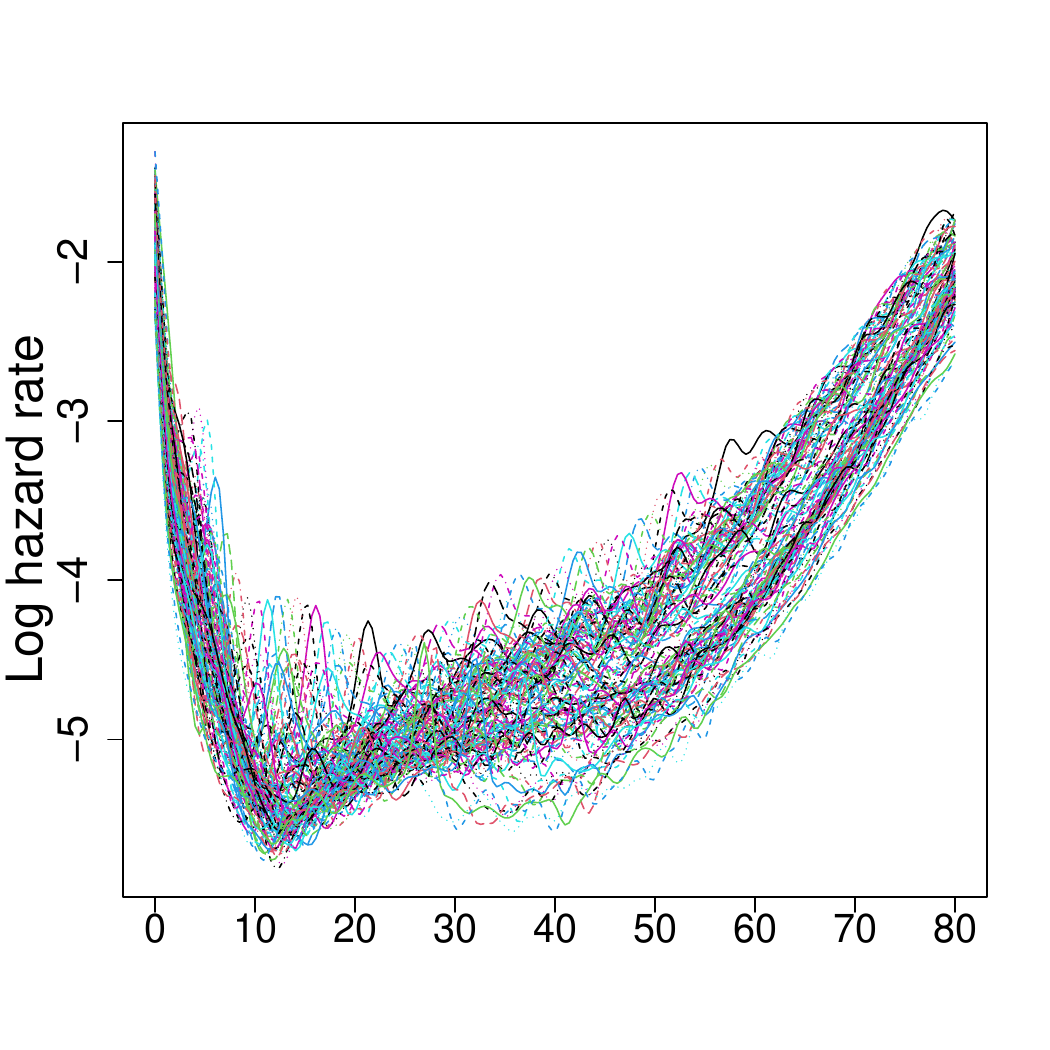}
	
	\caption{Log-hazard rates as a function of age for Swedish female cohorts born in the years 1751-1894.}
	\label{fig_datasw}
	
\end{figure}

The Swedish mortality data, available from the Human Mortality Database  (\url{http://mortality.org}), are regarded as a very reliable dataset on long-term longitudinal mortalities. In particular, we focus on the log-hazard rate functions of   the Swedish females mortality data for year-of-birth cohorts that  refer to females born in the years 1751-1894 with ages 0-80. The value of a log-hazard rate function at a given age is the natural logarithm of the ratio of the number of females who died at that age and the number of females alive with the same age. Note that, those  data have been analysed also by \cite{chiou2009modeling} and \cite{ramsay2009functional}. Figure \ref{fig_datasw} shows the 144 log-hazard functions.

The aim of the analysis is to explore the relationship of the log-hazard rate function for a given year with  the log-hazard rate curve of the previous year by means of the model reported in Equation \eqref{eq_lm}. Our interest is to identify what features of the log-hazard rate functions for a given year influence the log-hazard rate of the following year.

\begin{figure}
	
	\centering
	\includegraphics[width=\textwidth]{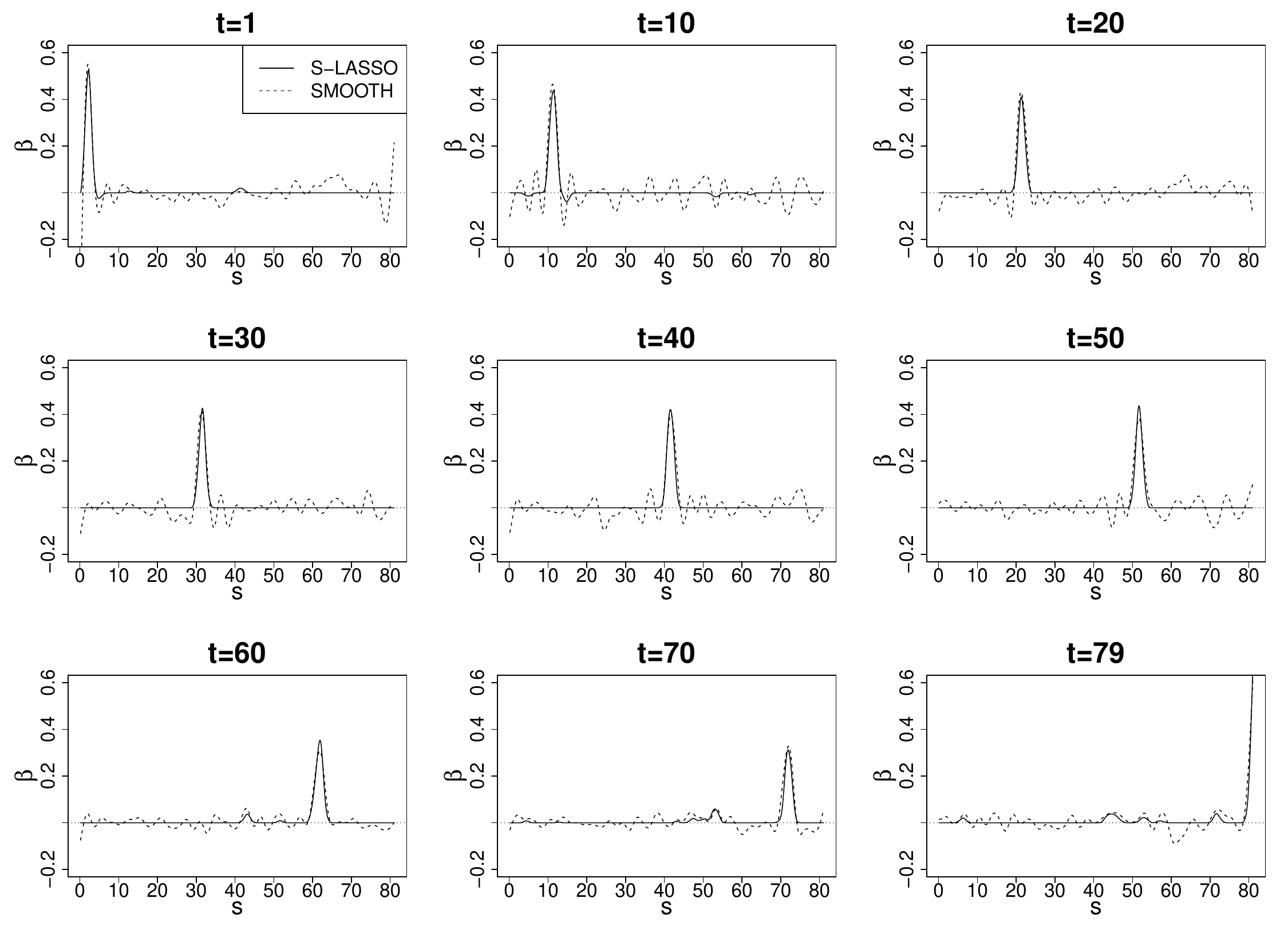}

	\caption{S-LASSO (\full) and SMOOTH (\dashed) estimates of the coefficient functions $\beta$  at different values of $t$ for the Swedish mortality data.}
	\label{fig_betasw}
	
\end{figure}

Figure  \ref{fig_betasw} shows the S-LASSO and SMOOTH estimates of  coefficient function $\beta$.
The unknown parameters to obtain  the SMOOTH and S-LASSO estimates are chosen as in the Canadian weather  example, but in this case   B-splines are used for both estimators.
The S-LASSO estimate is zero almost over all the domain except for few regions.
In particular, at given $t$, the S-LASSO estimate is different from zero  in an interval located right after  that age.
This can likely support the conjecture that if an event influences the mortality of the Swedish female  at a given age, it impacts on the the death rate below that age born in the following  years. 
Nevertheless, this expected dependence is poorly pointed out by the SMOOTH estimator, where this behaviour is  confounded by less meaningful periodic components.
It is  interesting to note that  the S-LASSO estimate at high values of $t$  is slightly different from zero  for ages ranging from 40 to 60. This shows that if an event affecting the death rate occurs in that range,  the log-hazard functions of the following cohorts will be influenced at high ages (i.e., corresponding to high values of $t$).
On the contrary, the wiggle of the SMOOTH estimate does not allow drawing such conclusions.

\begin{figure}
	
	\centering
	\begin{subfigure}[b]{0.45\textwidth}
		
		\centering
		\includegraphics[width=\textwidth]{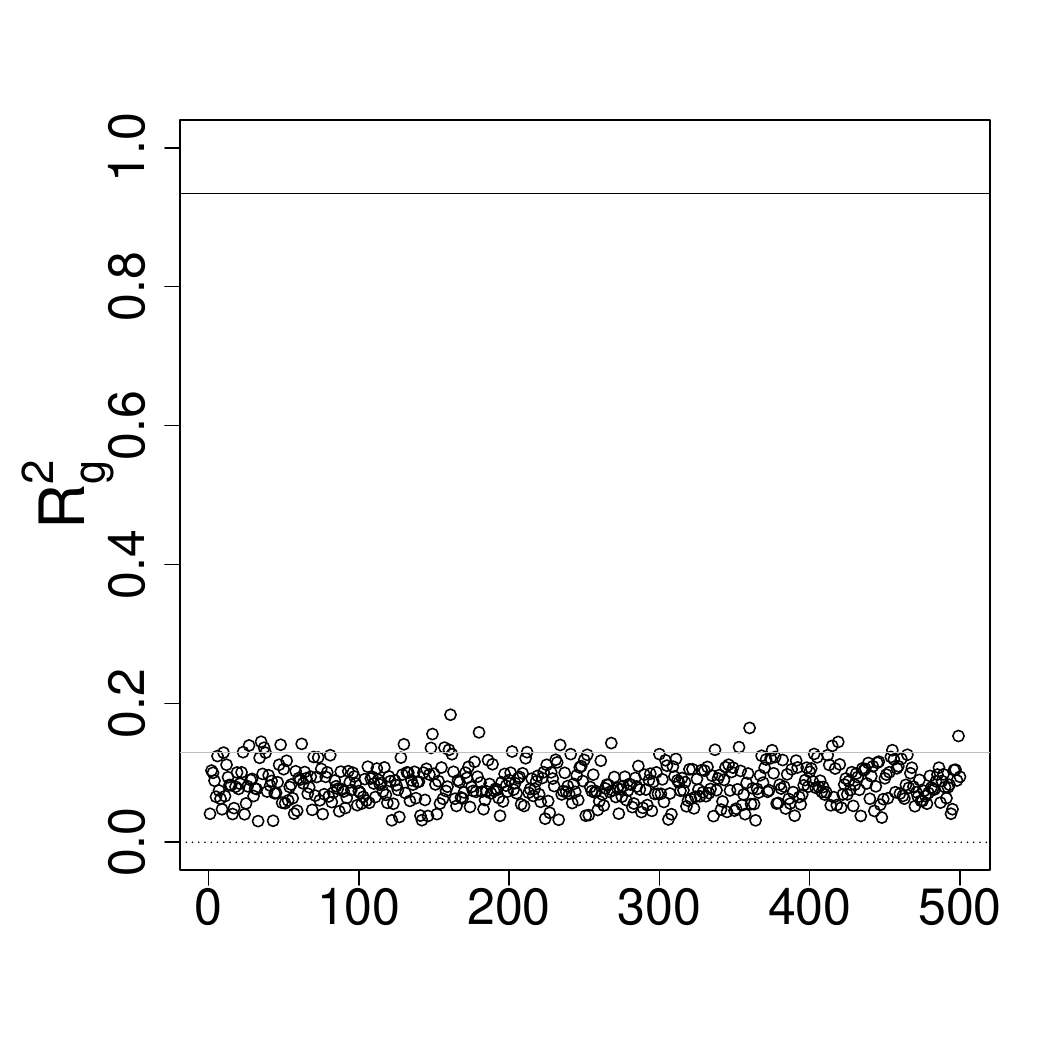}
		\vspace{-0.8cm}
		\caption{}
		\label{subfig_persw_1}
	\end{subfigure}
	\begin{subfigure}[b]{0.45\textwidth}
		
		\centering
		\includegraphics[width=\textwidth]{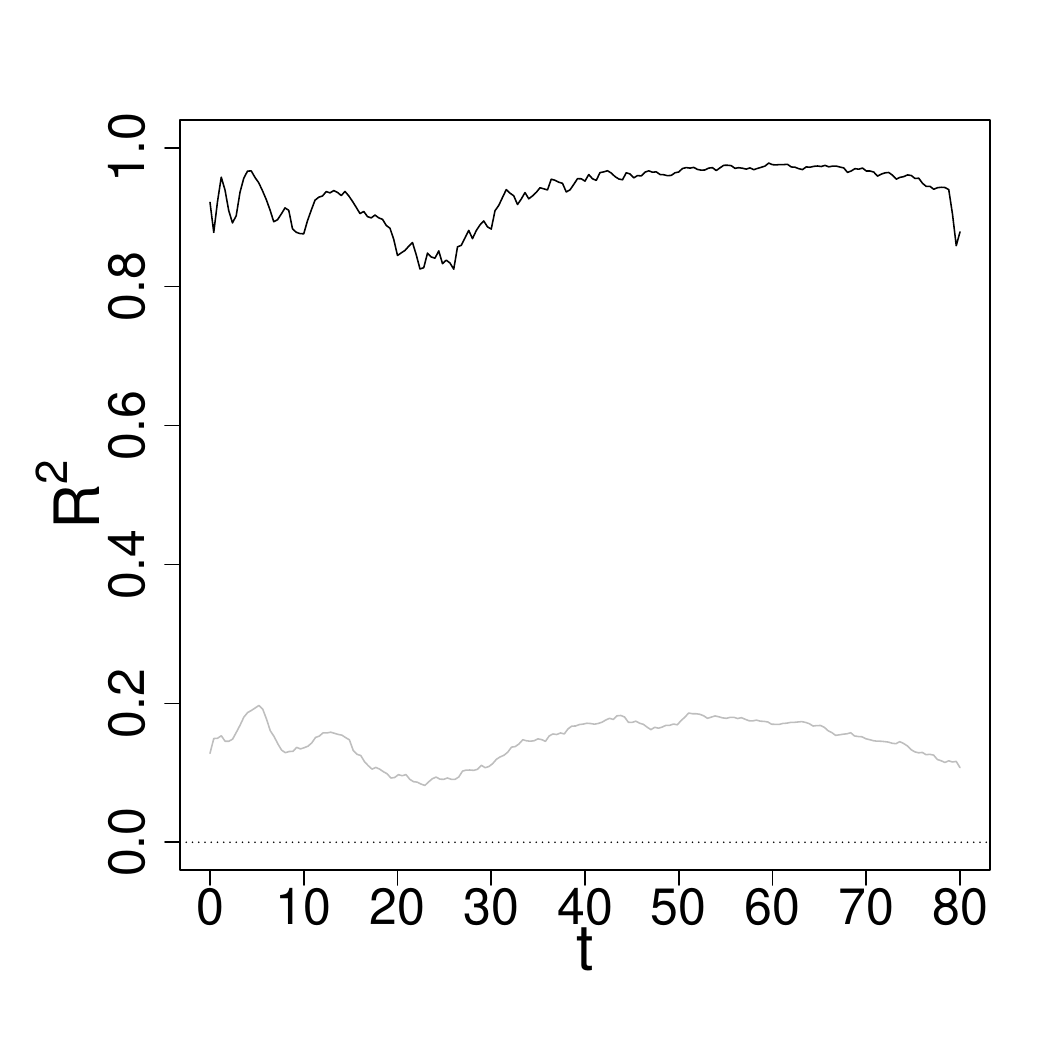}
		\vspace{-0.8cm}
		\caption{}
		\label{subfig_persw_2}
	\end{subfigure}

	\caption{For the Swedish mortality data, \subref{subfig_percan_1} $R^{2}_{g}$ from permuting the response 500 times, where the black line corresponds to the observed $R^{2}_{g}$ and the grey line  to  the  $95$th sample percentile; \subref{subfig_percan_2} the black line is the observed $R^{2}$ and the grey line is the pointwise $95$th sample percentile curve. }
	\label{fig_persw}
	
\end{figure}

Finally, we perform the two permutation tests already described  in the Canadian weather data example.
Figure \ref{fig_persw} shows the results. Both the observed  $R^{2}_{g}$  and  $R^{2}$ are far above the $95$th sample percentile (Figure \ref{subfig_persw_1})  and the pointwise $95$th sample percentile curve (Figure \ref{subfig_persw_2})  respectively. This significantly  evidences  a  relation between two consecutive log-hazard rate functions for all ages.
\subsection{Ship CO\textsubscript{2} Emission Data}
\label{sec_realgr}
\begin{figure}[H]
	
	\centering
	\includegraphics[width=0.45\textwidth]{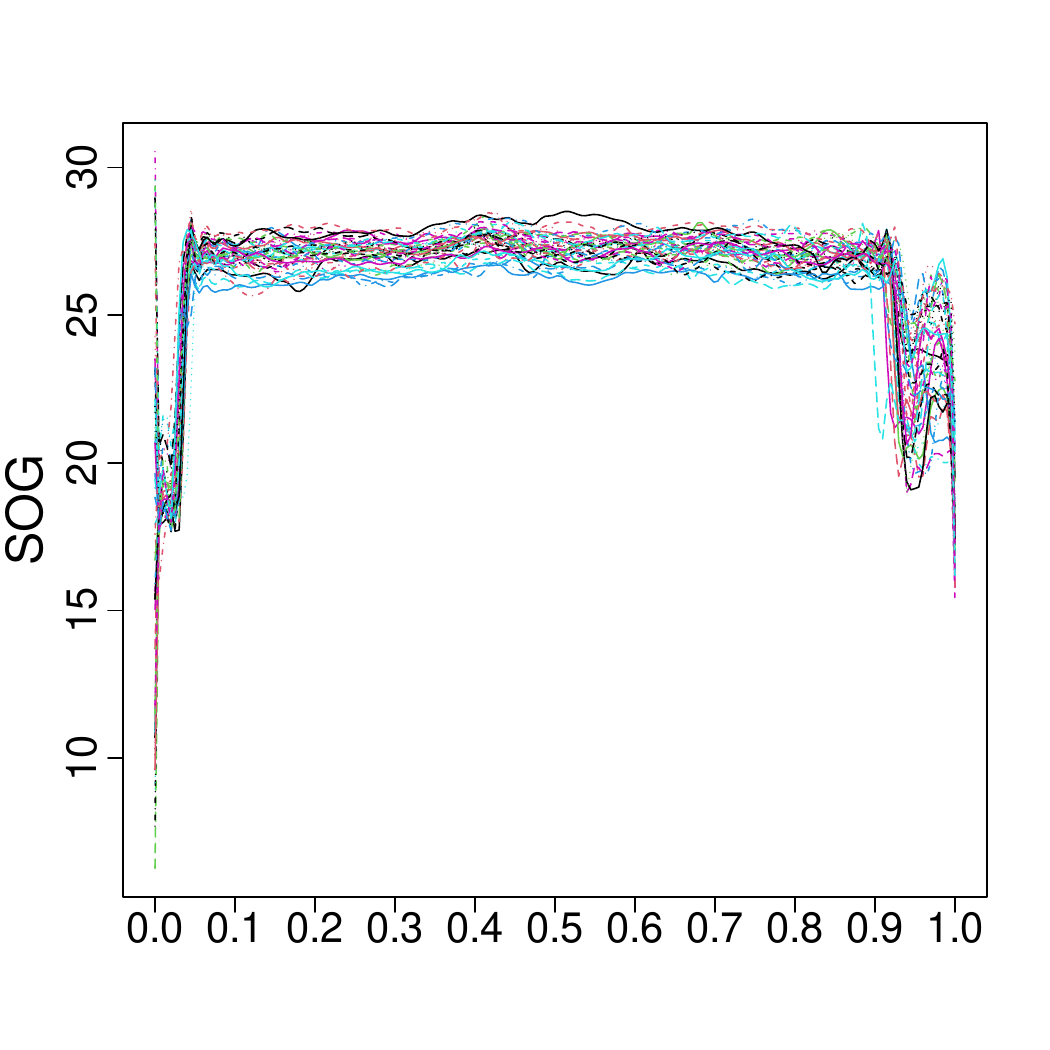}
	\includegraphics[width=0.45\textwidth]{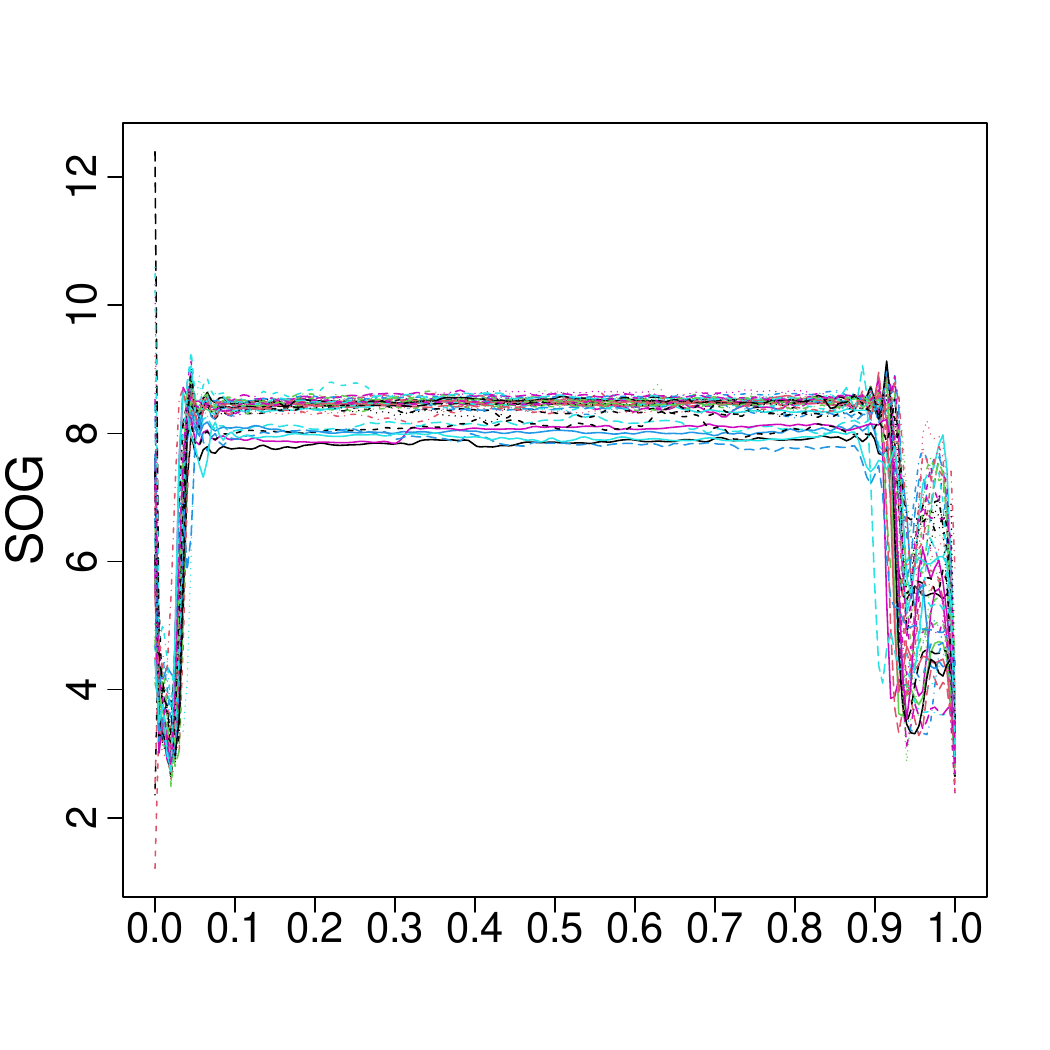}

	\caption{SOG and FCPH observations from a Ro-Pax ship.}
	\label{fig_datagrimald}
	
\end{figure}

\begin{figure}
	
	\centering
	\includegraphics[width=\textwidth]{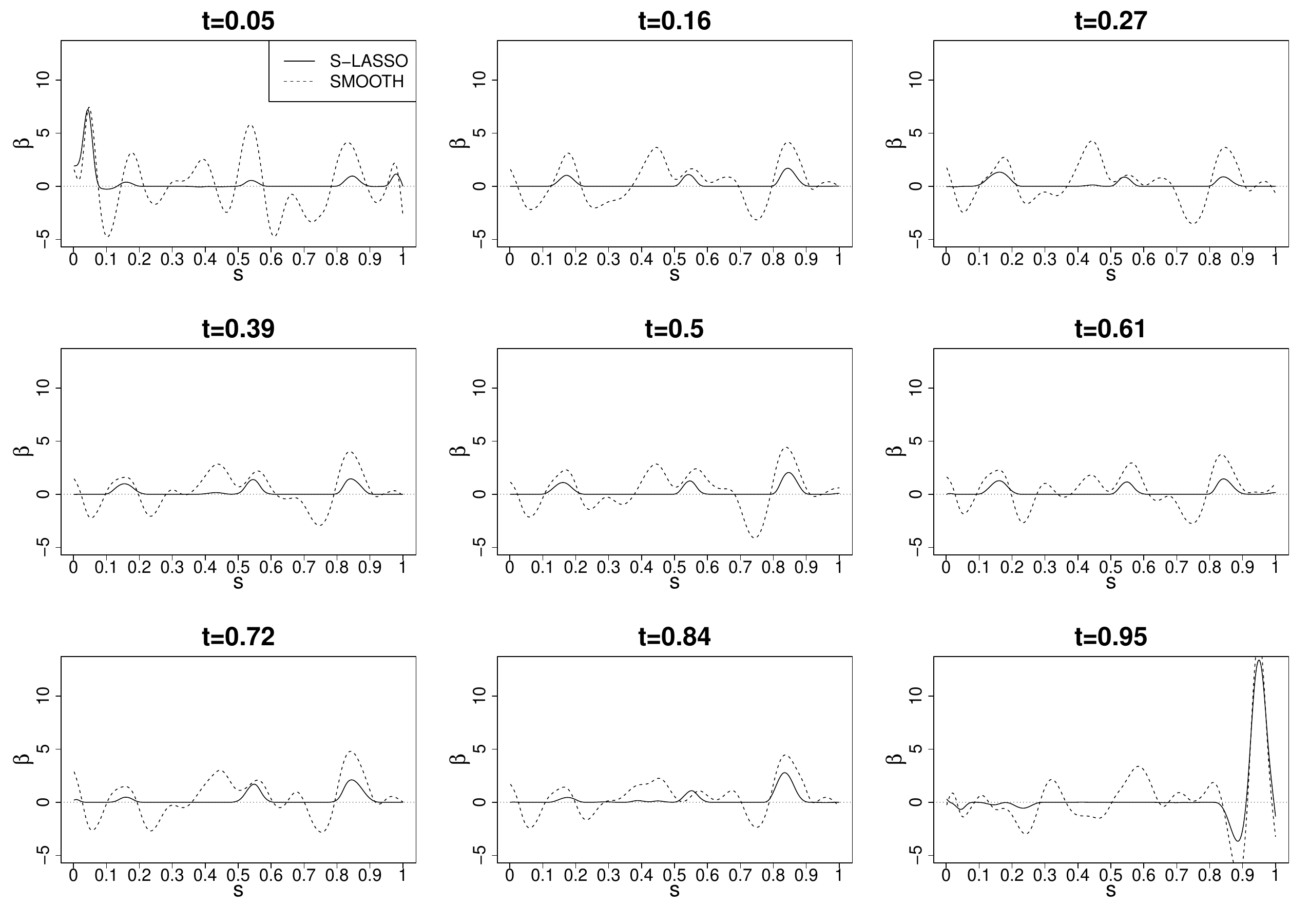}
	\caption{S-LASSO (\full) and SMOOTH (\dashed) estimates of the coefficient functions $\beta$  at different values of $t$ for the ship CO\textsubscript{2} emission data.}
	\label{fig_betagr}
	
\end{figure}

The ship CO\textsubscript{2} emission data have been previously studied  in  \citep{lepore2018analysis,reis2019predicting,capezza2020control,centofanti2020functional,centofanti2020adaptive}. The dataset, provided by  the shipping company Grimaldi Group, regards  the issue of monitoring fuel consumptions or CO\textsubscript{2} emissions for a Ro-Pax ship that sails along a route in the Mediterranean Sea.
The aim of the analysis is to study  the relation between the \textit{fuel consumption per hour} (FCPH), which is  a proxy of  the CO\textsubscript{2} emissions, and the \textit{speed over ground} (SOG), assumed as predictor. The observations considered were recorded from 2015 to 2017.
Figure \ref{fig_datagrimald} shows the 44 available observations of SOG and FCPH \citep{centofanti2020functional}.
Then, Figure  \ref{fig_betagr} displays the S-LASSO and SMOOTH estimates of  coefficient function $\beta$ estimated as described in the Swedish mortality example.
The S-LASSO coefficient function estimate also in this example is different from zero over a small portion of domain.
Specifically, the FCPH during the navigation phase (i.e., $ t\in \left[0.1,0.9\right] $) is positively influenced by the SOG,  in three specific voyage instants, viz., $ s\approx0.15,0.55,0.85 $, where the coefficient function estimate is positive. Thus, the FCPH during the navigation phase positively depends on the SOG at the start, in the middle and at the end of the navigation phase.
Differently, during  acceleration ($ t\in\left[0,0.1\right] $) and deceleration ($ t\in\left[0.9,1\right] $) phases, the relationship between the FCPH and the SOG is mainly concurrent. That is, the  FCPH observed at a given time instant is strictly related to the SOG observed at the same time.
On the contrary, such interpretation of the FCPH-SOG relationship is not easily obtained through the  analysis of the SMOOTH coefficient function estimate, which is overall different from zero.
Moreover, the S-LASSO and the SMOOTH estimates achieve 10-fold cross-validation mean squared errors of 0.077 and 0.093, respectively. Thus, the proposed estimator achieves slightly better prediction performance than the competitor.

\begin{figure}
	
	\centering
	\begin{subfigure}[b]{0.45\textwidth}
		
		\centering
		\includegraphics[width=\textwidth]{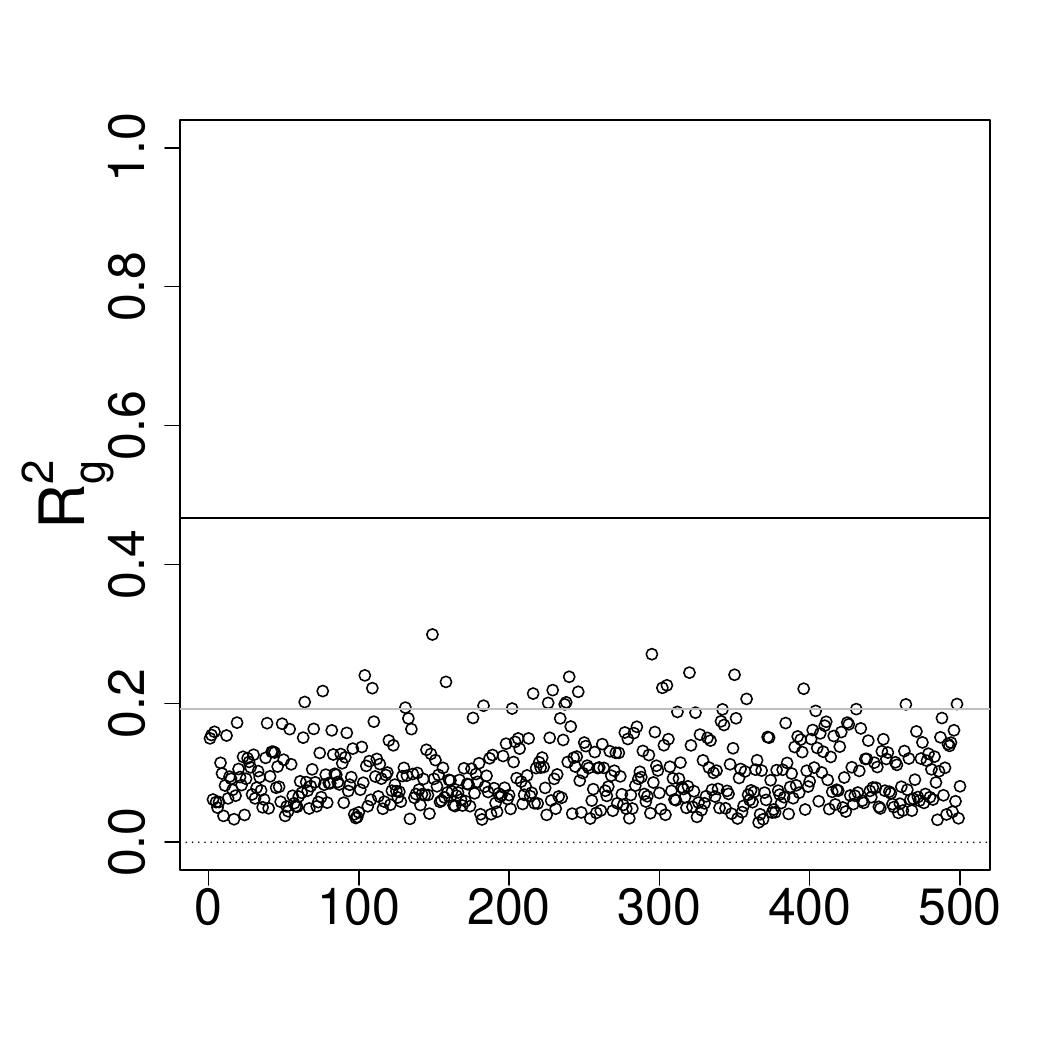}
		\vspace{-0.8cm}
		\caption{}
		\label{subfig_pergri_1}
	\end{subfigure}
	\begin{subfigure}[b]{0.45\textwidth}
		
		\centering
		\includegraphics[width=\textwidth]{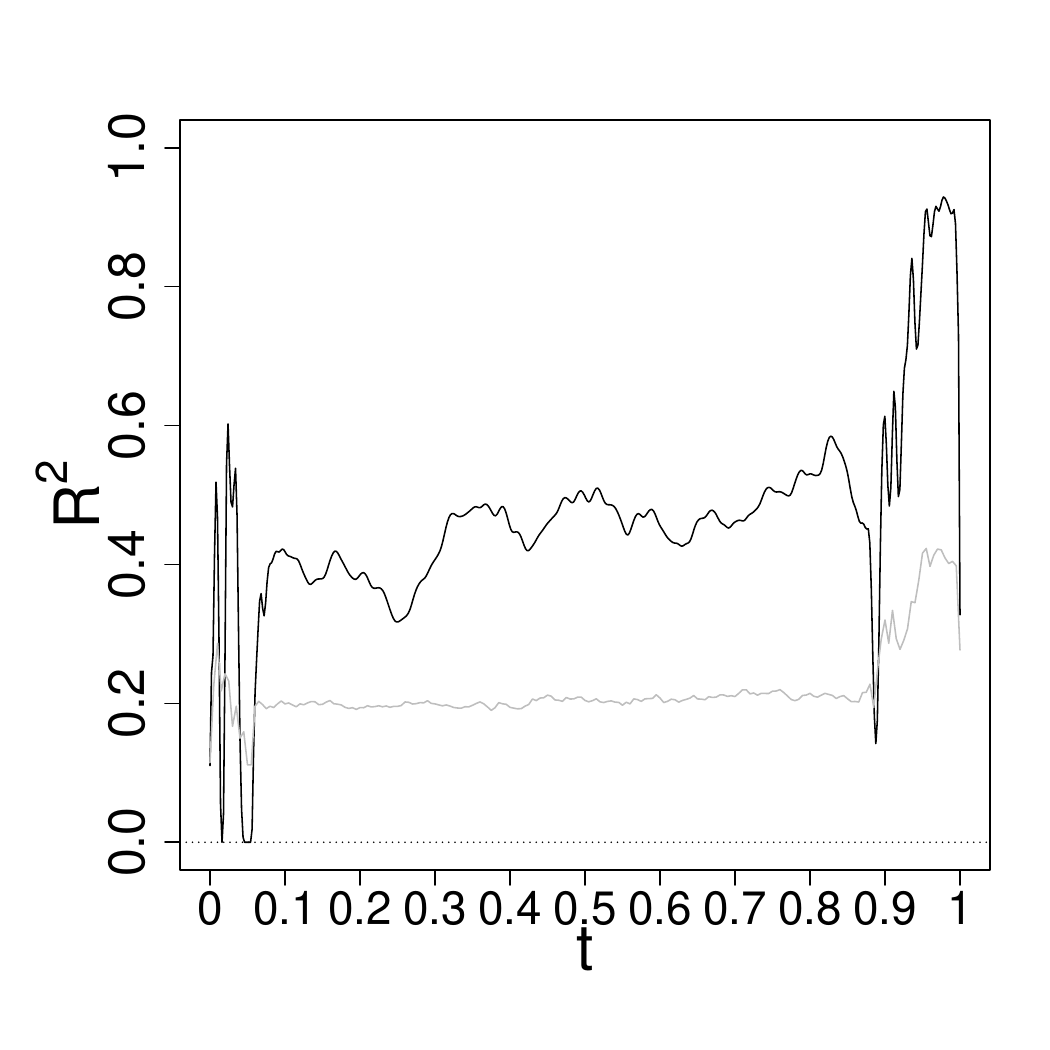}
		\vspace{-0.8cm}
		\caption{}
		\label{subfig_pergri_2}
	\end{subfigure}

	\caption{For the ship CO\textsubscript{2} data, \subref{subfig_pergri_1} $R^{2}_{g}$ from permuting the response 500 times, where the black line corresponds to the observed $R^{2}_{g}$ and the grey line  to  the  $95$th sample percentile; \subref{subfig_pergri_2} the black line is the observed $R^{2}$ and the grey line is the pointwise $95$th sample percentile curve. }
	\label{fig_pergri}
	
\end{figure}

Figure \ref{fig_pergri} shows the results for the two permutation tests already described  in the Canadian weather and Swedish mortality data examples. The relation between the FCPH and the SOG can be then considered as  significant,  as both the observed  $R^{2}_{g}$  and  $R^{2}$ are  above the $95$th sample percentile (Figure \ref{subfig_pergri_1})  and the pointwise $95$th sample percentile curve (Figure \ref{subfig_pergri_2}),  respectively. 

\newpage
\newpage
\section{Conclusion}
\label{sec_conc}

The LASSO  is one of the most used and popular method to estimate   coefficients in  classical linear regression models as it ensures both prediction accuracy and interpretability of the phenomenon under study (by simultaneously performing variable selection).
In this paper, we propose the S-LASSO estimator for the coefficient function of the Function-on-Function (FoF) linear regression model, which is an extension to the functional setting of the multivariate LASSO estimator.
As the latter, the S-LASSO estimator is  able   to increase  both the prediction accuracy of the estimated model, via continuous shrinking,   and the interpretability, by identifying the null region of the regression coefficient, i.e., the region where the coefficient function is exactly zero. 

The  S-LASSO estimator  is obtained by combining  several elements: the  \textit{functional LASSO penalty} (FLP), which has the task of shrinking towards zero the  estimator on the null region; the B-splines, which are essential to ensure sparsity of the estimator because of the compact support property; and two roughness penalties, which are needed to ensure smoothness of the estimator on the non-null region, also when the number of B-splines escalates.
We  proved that the S-LASSO estimator is both estimation  and point-wise sign consistent, i.e.,  the estimation error in terms of $L^{2}$-norm goes to zero in probability and  the S-LASSO estimator has the same sign of the true coefficient function with probability tending to one.
Moreover, we  showed via an extensive Monte Carlo simulation study that, with respect to other methods that have already appeared in the literature, the S-LASSO estimator is much more interpretable, on the one hand,  and  has still  good estimation and appealing predictive performance, on the other.
However, consistently with  the behaviour of the classical LASSO estimator \citep{fan2004nonconcave}, the S-LASSO  estimator is found sometimes  to  over-shrink the coefficient function  over the non-null region.

Note that the S-LASSO method is in the spirit of the \textit{fused LASSO} of \cite{tibshirani2005sparsity} in the multivariate context. Both methods rely on  penalties that   encourage sparsity in the coefficients and impose smoothness in the coefficient profile. The smoothness penalties of the  proposed S-LASSO estimator and the fused LASSO approach have a  different nature. In the fused LASSO,  the differences of two consecutive coefficients are in fact penalized in their absolute value thus  shrinking consecutive coefficients to the same value. On the contrary, the S-LASSO estimator relies on quadratic smoothness penalties that do not enjoy the sparseness property. Moreover, while the fused LASSO considers penalization of the first differences, the S-LASSO estimator allows the penalization of several types of smoothness, through the $m$s-th and $m$t-th order linear differential operators applied to the coefficient function.

To the best of the authors knowledge, this is the first work that addresses  the issue of  interpretability, intended as sparseness of the coefficient function,  in the  FoF linear regression setting. However, although the FLP produces an estimator with good properties,  other  penalties, e.g. the \textit{SCAD} \citep{fan2001variable} and \textit{adaptive LASSO} \citep{zou2006adaptive}, properly adapted to the functional setting, may guarantee  even better performance both in terms of interpretabilty and prediction accuracy, and are, 
indeed,  subjects of ongoing research.

\bibliography{References,references2}

\end{document}